\documentclass[sigconf, nonacm]{acmart}
\usepackage{makecell} %
\usepackage{graphicx} %
\usepackage{amsmath, amsthm}
\usepackage{booktabs}
\usepackage{caption}
\usepackage{subcaption}
\usepackage{enumitem}
\usepackage{algorithm}
\usepackage{algorithmic}
\usepackage{hyperref}
\usepackage{inputenc}
\usepackage{xcolor}
\usepackage{geometry}
\AtBeginDocument{%
  }

\setcopyright{none} %

\copyrightyear{2025}
\acmYear{2025}
\acmDOI{}
\acmConference[Preprint.]{}{2025}{}
\acmISBN{}

\begin{document}

\title[Superhuman Game AI Disclosure]{Superhuman Game AI Disclosure: Expertise and Context Moderate Effects on Trust and Fairness}

\author{Jaymari Chua}
\affiliation{%
  \institution{CSIRO's Data61 and UNSW}
  \city{Sydney}
  \state{NSW}
  \country{Australia}
}

\author{Chen Wang}
\affiliation{%
  \institution{CSIRO's Data61}
  \city{Sydney}
  \state{NSW}
  \country{Australia}
}

\author{Lina Yao}
\affiliation{%
  \institution{CSIRO's Data61 and UNSW}
  \city{Sydney}
  \state{NSW}
  \country{Australia}
}

\begin{abstract}
As artificial intelligence surpasses human performance in select tasks, disclosing superhuman capabilities poses distinct challenges for fairness, accountability, and trust. However, the impact of such disclosures on diverse user attitudes and behaviors remains unclear, particularly concerning potential negative reactions like discouragement or overreliance. This paper investigates these effects by utilizing \emph{Persona Cards}: a validated, standardized set of synthetic personas designed to simulate diverse user reactions and fairness perspectives. We conducted an ethics board-approved study (N=32), utilizing these personas to investigate how capability disclosure influenced behaviors with a superhuman game AI in competitive StarCraft~II scenarios. Our results reveal transparency is double-edged: while disclosure could alleviate suspicion, it also provoked frustration and strategic defeatism among novices in cooperative scenarios, as well as overreliance in competitive contexts. Experienced and competitive players interpreted disclosure as confirmation of an unbeatable opponent, shifting to suboptimal goals. We release the Persona Cards Dataset, including profiles, prompts, interaction logs, and protocols, to foster reproducible research into human alignment AI design. This work demonstrates that transparency is not a cure-all; successfully leveraging disclosure to enhance trust and accountability requires careful tailoring to user characteristics, domain norms, and specific fairness objectives.
\end{abstract}



\received{15 January 2025}
\received[revised]{2 April 2025}

\maketitle

\section{INTRODUCTION} \label{sec:introduction}

AI systems operate behind the scenes in domains such as clinical decision support, e-commerce, and multiplayer gaming, prompting critical questions about fairness, accountability, and transparency \cite{friedler2021possibility, pagano2023bias}. An influential example is DeepMind's AlphaStar, which achieves pro level play in the real-time strategy (RTS) game StarCraft~II through inhumanly fast maneuvers and near-perfect resource management \cite{silver2017mastering, vinyals2019grandmaster}. Although these achievements highlight significant AI capabilities, they also raise concerns about how humans perceive and place trust in AI with a seemingly insurmountable machine advantage described as \emph{superhuman}.

In StarCraft~II and other multiplayer gaming contexts, much research has addressed overt harassment and toxic gaming behavior \cite{laato2024traumatizing, kou2020toxicity}, yet there is insufficient understanding of how subtler manifestations of AI dominance affect user trust and fairness perceptions particularly when systems like AlphaStar deploy superhuman capabilities \cite{vinyals2019grandmaster}. This gap becomes more critical as transparency about AI advantages can simultaneously mitigate suspicion or provoke frustration \cite{, eslami2016user}, depending on whether users see these capabilities as rightful skill or an unfair edge. Transparency issues also arise in assistant scenarios where AI serves as an ally, as even in cooperative settings, the application of advanced AI presents challenges. Empirical evidence from customer service suggests that explicit disclosure of AI utilization evoke perceptions of manipulation \cite{seeger2021texting, cai2019hello, gillespie2018custodians, van2024understanding, adam2021ai}. In recommendation systems, despite its intended helpfulness, highly accurate predictions and personalized outputs may be construed as a violation of privacy \cite{yang2023dynamic, zhu2022privacy, logg2019algorithm}.
Despite ongoing efforts to improve AI transparency, it remains unclear whether disclosing \emph{superhuman AI} model capabilities may either foster trust calibration or confirm doubts about cheating and undue influence \cite{holstein2022designing}. Even with carefully framed disclosures, transparency alone does not guarantee optimal trust  calibration~\cite{jiang2024self,zhang2024sa,hagendorff2023human}. AI explanations confuse users if explanations exceed their expertise ~\cite{bansal2021does}; and novice players might overly trust  a system perceived as infallible leading to AI overreliance ~\cite{kocielnik2019will,schmitt2024role}. Consequently, an investigation into how model capability disclosure affects user expertise and domain context is important for achieving ethical and responsible AI deployment. We do not yet understand how explicit disclosure of superhuman capabilities shapes user attitudes and behaviors across adversarial versus cooperative settings, for users with different expertise levels.

To address the laid out challenges posed by disclosing \emph{superhuman AI} capabilities, our research investigates the following question: \textbf{RQ1:} \emph{How does concealment versus disclosing superhuman AI capabilities influence user perceptions of fairness, toxicity, and trust?} \textbf{RQ2:} \emph{And to what extent do user characteristics attributes to personality or notions of play style, particularly expertise level, moderate the impact of superhuman AI disclosure on user attitudes and behaviors?} 

Naturally suited for investigating our research questions is StarCraft II, a real-time strategy (RTS) game known to reliably elicit complex strategic behavior and emotional investment, making it particularly effective for exploring trust calibration and fairness perceptions under superhuman AI contexts. However, despite the carefully framed disclosures suggested by prior research, transparency alone may fail to consistently optimize trust calibration due to varying user expertise and predispositions toward technology. Our study hypothesizes that explicit disclosure of superhuman AI capabilities may heighten user frustration and suspicion, particularly among novices, whereas experts may respond with suboptimal strategy adaptations.

Our research contribution to responsible AI involves developing a concept called \emph{Persona Cards}, inspired by model cards. These cards serve as a tool to track user values and biases that influence ratings, enabling the creation of synthetic human participant data. The goal is to model responses from casual to mid-level and expert synthetic persona players, varying whether they are informed about the AI’s superhuman capabilities. This allows us to address our first research question: how do different levels of disclosure about the AI’s capabilities affect user reactions? As for our second research question, each synthetic persona incorporates defined cognitive architectures and belief updating mechanisms. This enables us to examine how various factors, such as different levels of openness and acceptance of technology among different segments of the population, and prior assumptions, shape reactions to superior AI performance. To further validate our data collection, we analyse quantitative telemetry (APM logs and resource utilisation) and post-match interviews to ground the synthetic data generation process and assess fairness, enjoyment, and frustration. We tracked user reactions and perceptions through persona-based data generation, incorporating detailed cognitive architecture attributes.

In order to address ecological validity and ethical considerations regarding synthetic persona data, we conducted an IRB approved experiment involving real human participants (N=32) playing against our pro-level StarCraft II game AI with variable APM rates. Participants were randomly assigned to either \emph{disclosure} or \emph{non-disclosure} conditions, ensuring that our hypothesis remained concealed while still informing them of the model’s advanced capabilities in the disclosure arm. Immediately after each match, we debriefed participants to mitigate potential frustration.

Our findings show that disclosure significantly shapes perceptions, but its effects vary on the user’s expertise and domain. In StarCraft~II, novice players sometimes developed resentment or excessive reliance, expecting flawless performance. On the other hand, expert players who usually caution against superhuman play and experimented with novel strategies that could potentially outperform the AI, after being informed of the AI’s superhuman capabilities, concluded formidable gameplay as unbeatable. This highlights that humans are prone to be misled to pursue subpar goals, such as mini-games to prolong the game timer before defeat from the AI, instead of investing their time and effort in achieving their original and true objective to genuinely beat the AI. The results matched with the IRB approved study, as human participants also excessively relied on the system based on the researchers’ disclosure statement.

Alongside findings, we introduce \emph{Persona Cards~Dataset}, a corpus of systematically generated persona interactions grounded in cognitive architectures, offering high reproducibility and controlled variations. Analyzing how superhuman advantages influence fairness, toxicity, and trust, \emph{Persona Cards~Dataset} addresses both the ethical challenges and the methodological demands of studying advanced AI across different tasks without confounding biases. Therefore, our study offers the following contributions to advance the responsible development and deployment of AI systems:
\begin{itemize}
    \item \emph{Methodological Contribution}. We introduce \emph{Persona Cards}, a novel systematic framework explicitly documenting user values and biases that inform rating behaviors, enabling more transparent, accountable, and fairness-aware synthetic persona generation for AI assessments.
    \item \emph{Empirical Contribution}. We empirically demonstrate that disclosure regarding an AI’s superhuman capabilities move fairness judgments validated across both adversarial and collaborative contexts.
    \item \emph{Dataset Contribution}. We contribute \emph{Persona Cards~Dataset}, an open source repository of persona card interactions grounded within defined cognitive architectures, enabling reproducible experimentation and cross-domain insights.
    \item \emph{Practical Design Guidelines}. We propose concrete adaptive design strategies, emphasizing balanced transparency tailored to user expertise, to practically foster equitable AI experiences and strengthen real-world accountability and fairness standards.
\end{itemize}

\hyperref[sec:related-work]{Section~\ref{sec:related-work}} reviews perceptions of fairness, toxicity in gaming, and disclosure as a mechanism for establishing trust. \hyperref[sec:methods]{Section~\ref{sec:methods}} details synthetic persona as a method for HCI research attribute manipulation. \hyperref[sec:results]{Section~\ref{sec:results}} presents the empirical findings, including statistical confirmation. \hyperref[sec:discussion]{Section~\ref{sec:add_results}} interprets these results for fairness and trust calibration. \hyperref[sec:discussion]{Section~\ref{sec:discussion}} concludes, and \hyperref[sec:ethics]{Section~\ref{sec:ethics}} reflects on ethics and limitations of our study.

\section{RELATED WORK} \label{sec:related-work}

\subsection{Socio-Technical Governance and AI Policy}
There is a movement of artificial intelligence technology that is beyond technical considerations towards robust socio-technical governance frameworks and effective AI policy. As AI systems become increasingly embedded in societal structures, there is a clear consensus that ensuring responsible development and deployment requires bridging the gap between high-level ethical principles and concrete, enforceable mechanisms. \cite{shneiderman2020bridging} articulate this shift, arguing that abstract values like fairness, transparency, and accountability must be operationalized through specific laws, regulations.

A central theme in contemporary scholarship is the framing of AI governance as an inherently socio-technical challenge. This perspective in evaluating AI systems in isolation, instead considering the complex interplay between the technology, its developers, deploying organizations, end-users, and regulatory bodies \cite{selbst2019fairness}. Issues such as bias, safety, or trustworthiness are understood not merely as technical properties but as emergent outcomes of this broader ecosystem. \cite{manzini2024should}, for example, propose that establishing trust in advanced AI assistants necessitates evaluating evidence of competence and alignment across multiple levels: the AI system's design features, the organizational practices of its creators (including ethics reviews and testing protocols), and the effectiveness of third-party governance mechanisms like external audits or certifications. This multi-level, contextual approach, examining AI within its intricate web of human and institutional interactions, is increasingly recognized as essential for effective governance \cite{selbst2019fairness}. Addressing the unique challenges posed by advanced AI systems, such as large language models and autonomous agents, has become a focal point for policy and research. Concerns regarding misinformation, market concentration, and potential risks have spurred the development of specific regulatory frameworks. Notable examples include the US NIST AI Risk Management Framework \cite{ai2023artificial} and the EU AI Act \cite{nikolinakos2023eu}, both advocating for risk-based approaches that impose stricter requirements on high-impact applications. 

\subsection{Perceived Unfairness of Game AI Capability}
Fairness and toxicity have been extensively studied in human-to-human settings, focusing on verbal harassment, abuse, or trolling~\cite{kuznekoff2013content,kou2020toxicity,laato2024traumatizing}. However, AI opponents introduce new dimensions of mechanical or strategic toxicity. In real-time strategy (RTS) games such as StarCraft II and Dota 2, AI systems frequently exceed professional human norms of around 300 actions per minute (APM)~\cite{vinyals2019grandmaster}, prompting accusations of inhuman or unfair play~\cite{peters2020characterizing}. This is consistent with research on learning from human pro play in StarCraft~II, where high APM in AI players was perceived as unfair. Developers sometimes cap AI actions, as in limiting AlphaStar's APM, to avoid overwhelming players~\cite{vinyals2019grandmaster,berner2019dota}, yet these caps rarely address whether explicitly stating "this game AI is superhuman" shifts player perceptions of cheating~\cite{pfau2024damage, kim2024toxicity, kocielnik2019will}. The potential for AI to exploit game mechanics, often referred to as reward hacking, can further exacerbate perceptions of unfairness \cite{amodei2016concrete,krakovna2020specification}.
Reward hacking can also yield seemingly illegitimate strategies that exploit the game's reward function~\cite{amodei2016concrete,krakovna2020specification}, compounding notions of unfair advantage. A previous study has shown that toxicity arises from multiple factors, including competitive mechanics and perceived misalignment with social norms \cite{beres2021don}. Our research extends this by investigating how labeling an AI as superhuman might exacerbate negativity if players interpret any deviation from human-like performance as cheating. These concerns parallel the alignment problem~\cite{hadfield2017inverse,lehman2018surprising}, wherein AI systems pursue narrow objectives that may conflict with user expectations of fair or balanced gameplay.
Henceforth, our work builds on these insights by examining how superhuman disclosure interacts with these established challenges in competitive gaming AI, and takes inspiration from work on how the relation to human competence relates to the reliance on AI \cite{he2023knowing}.

\subsection{Effects of Disclosure on Trust}
Research on AI disclaimers spans a variety of interaction modes, often focusing on large language models (LLMs) \cite{gillespie2018custodians, jiang2024self}. While disclosing “This is AI generated” can help users calibrate trust, it does not consistently alter perceptions of toxicity \cite{zhang2024sa}. Studies on blame attribution for algorithmic harm suggest that user reactions depend on factors such as perceived fairness, the harmfulness of the AI’s actions, and personal attitudes toward assigning blame to machines \cite{lima2023blaming}. 
Complicating matters, LLMs may self-moderate upon realizing they are under scrutiny, reflecting moral judgments embedded in their training data \cite{ganguli2022red}.
Moreover, demonstrating trust on through complying with AI systems is not dictated solely by the accuracy or precision of the text output. For instance, anthropomorphic design cues have been shown to increase compliance with chat bot requests, yet explicit AI labeling can provoke skepticism or diminish persuasion if users feel misled \cite{adam2021ai}. Related work on avatar disclosure indicates that revealing an avatar’s artificial nature can influence how intensely participants play or engage, especially when the avatar’s appearance suggests human likeness \cite{visser2024impact}. However, in customer-service environments, disclosure of a chat bots artificial identity did not necessarily undermine social presence or reduce helpfulness, as users often prioritized perceived usefulness over identity status \cite{van2024understanding}.
User familiarity with AI also plays a significant role: individuals less experienced with AI may misjudge or overestimate toxicity in AI-generated content, while more experienced users are often unmoved by disclaimers \cite{hagendorff2023human}. In contrast, other studies indicate that disclaimers can heighten user caution and prevent AI overreliance on the systems \cite{wischnewski2023measuring, jiang2024self}.

\subsection{Conceptual Framework and Key Definitions}
\label{sec:conceptual_framework}
This scopes how we define and use the primary constructs examined in: \emph{toxicity}, \emph{fairness}, and \emph{trust}. Although these terms appear throughout multiple literature, we adopt specific working definitions tailored to our experimental contexts. We define \emph{toxicity} including toxic gaming behavior and provoking communication (biased or insulting responses) whether automated or user-driven that is perceived as harmful, abusive, or excessive and therefore annoying \cite{kou2020toxicity, laato2024traumatizing}. In our StarCraft II scenario, toxicity may take the form of in-game chat or even gameplay actions that implies unfair superiority or demeans the opponent. Our measurement focuses on \emph{perceived} negativity rather than explicit harassment alone. As for \emph{Fairness}, building on established notions in social and algorithmic contexts \cite{selbst2019fairness, jacovi2021formalizing}, we frame \emph{fairness} as the extent to which an AI operates within recognized boundaries of normative behavior, abiding by rules, community standards, and user expectations. In StarCraft II, fairness specifically entails: (1) adherence to typical mechanical and strategic limitations, (2) avoidance of “unseen” or privileged information not accessible to a human opponent, and (3) reliance on tactics that conform to community norms (e.g., not exploiting known design loopholes). Recent work further highlights the necessity for AI systems to demonstrate a perceived absence of bias or undue manipulation in textual, visual, or interactive outputs (and actions, in our case).

A key assumption in our study regarding fairness is that experienced players, given their higher socio-technical technology acceptance, may both recognize and tolerate more advanced AI capabilities, provided those capabilities remain within what they perceive as “fair play.” And for building on the concept of \emph{trust} in human–AI interaction \cite{bansal2021does} through two main indicators: the persona’s willingness to rely on the AI’s outputs (or play against it again, in the game scenario) and the degree of confidence that the AI is acting consistently with its disclosed capabilities. High trust emerges when the persona perceives the AI as reliable, transparent, and aligned with their expectations.

\paragraph{Placement in the Literature}

Prior research emphasizes the importance of operationalizing transparency, fairness, and accountability into enforceable socio-technical governance mechanisms and actionable policy frameworks~\cite{shneiderman2020bridging, selbst2019fairness, nikolinakos2023eu, ai2023artificial}. Yet, existing approaches largely presume AI systems operate at or near human competence, leaving a notable gap when dealing explicitly with superhuman capabilities~\cite{vaccaro2024combinations, schmitt_etal_2024}. This introduces novel complexities, particularly around fairness defined as adherence to community norms and ethical expectations~\cite{jacovi2021formalizing} and trust calibration, aligning users’ expectations accurately with the system’s true capabilities~\cite{lee2004trust, bansal2021does}. The following Methods section details our empirical approach to answering these targeted research questions comprehensively.

\section{METHODS} \label{sec:methods}

To investigate how transparency on AI capabilities influences user perceptions, we conducted experiments manipulating disclosures about AI skills within both competitive gaming (specifically StarCraft~II \cite{vinyals2019grandmaster, kim2024toxicity}) and cooperative chatbot interactions \cite{van2024understanding}. Informed by initial pilot studies and insights from persona simulations which indicated substantial overlap across various preliminary disclosure conditions (e.g., varying ability types, superiority levels, explanatory depth), our final experimental design employed a targeted set of disclosure conditions consistently applicable across both contexts. These conditions systematically varied key dimensions of the information revealed about the AI, allowing for comparison of transparency effects across distinct interaction paradigms for our ethics approved user study with human participants against superhuman game AI in StarCraft~II.

\subsection{Procedure Conditions}
\label{sec:procedure_conditions}

\begin{figure*}[htbp]
\footnotesize
    \centering
    \includegraphics[width=1\linewidth, page=1]{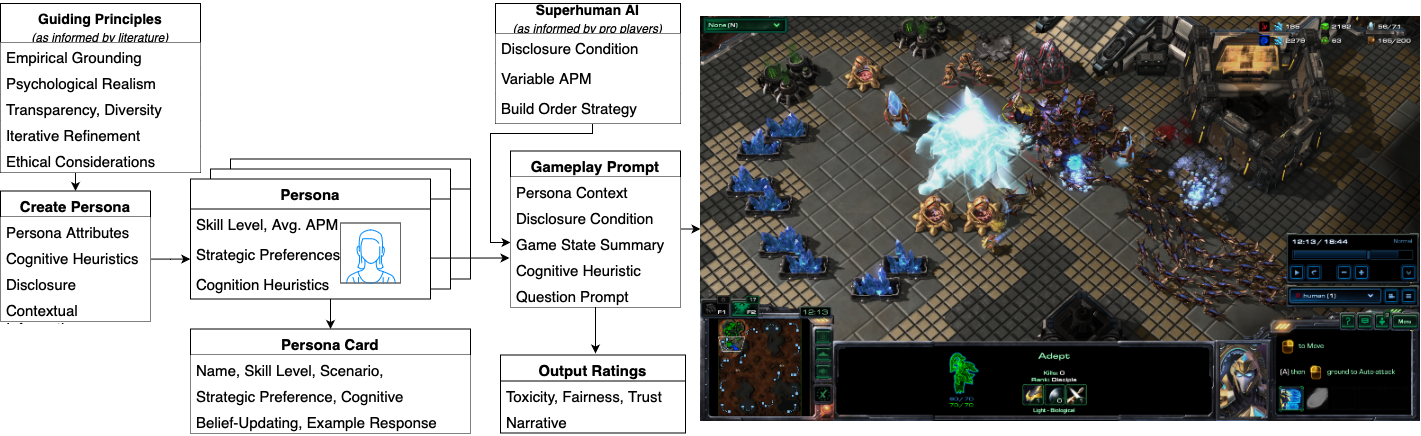}
    \caption{Framework for persona disclosure in Superhuman Game AI design: integrating empirical grounding, psychological realism, transparency, diversity, iterative refinement, and ethical considerations to generate contextually accurate gameplay personas and prompts, evaluated across toxicity, fairness, trust, and narrative dimensions.}
    \label{fig:procedure-diagram}
\end{figure*}

Answering \textbf{RQ1} means experiments on undisclosed versus disclosed superhuman skill and that should clarify on how persona beliefs were primed regarding the AI capabilities. Therefore, we implemented a between-subjects design, randomly assigning synthetic personas to one of several disclosure conditions. Each condition specified whether the AI was purportedly “human-level” or “superhuman,” and whether the persona received an explicit statement about the AI’s enhanced abilities. In practice, the \emph{no disclosure} condition did not reveal any superhuman trait, whereas the \emph{superhuman disclosure} condition explicitly stated, for instance, that the system had “far higher Actions Per Minute” or “markedly superior reasoning” compared to typical humans. The persona’s comprehension check ensured it internalized these statements accordingly. We drew on established protocols for controlling synthetic persona prior beliefs in experimental manipulations on trust \cite{kocielnik2019will, liu2022trustworthy}.

Each experiment began with an introduction to the persona and was briefed on and outlined the context: competitive behaviors bordering oppressive and toxic by the superhuman AI in StarCraft~II, and the disclosure statements were provided or withheld details about superhuman attributes according to the assigned condition. The persona’s internal belief model was then updated to reflect any disclosed capabilities. Next, the persona engaged in the main task. In any case, the AI behaved in ways calibrated to the assigned disclosure condition. Afterward, the persona completed a post-interaction questionnaire that measured perceived \emph{Toxicity}, \emph{Fairness}, and \emph{Trust}, each defined below and recorded on a 5-point Likert scale \cite{starke2021measuring, kocielnik2019will}. The persona also provided an open-text justification of its responses, thereby revealing how its internal heuristics framed the AI’s behavior. 
Personas received a brief scenario description, again either revealing or withholding superhuman attributes of the LLM. They then posed a standardized set of questions (factual, ethical, creative), which the LLM answered with varying degrees of correctness and occasional subtle toxicity \cite{datta2016algorithmic}. Specific triggers inserted ambiguous or biased statements, reflecting real-world vulnerabilities in generative models. When the superhuman condition was disclosed, the LLM sometimes reminded the persona of its own “advanced reasoning” or “vast knowledge,” thus providing a textual parallel to the in-game advantage references in StarCraft~II.

The resulting persona (documented via a card) specifies skill level, strategic preferences, and cognitive heuristics, and is then embedded into a gameplay simulation alongside a ``Superhuman AI'' configuration. The system is designed for flexible interchange, enabling real participants or an external game engine to replace any component, and allowing large language models to interface with or substitute modules as needed. Our recommendation is to use the personas to seed and inform an experiment with an advanced AI, and then proceeding with an ethical human study for grounding.

Synthetic personas seed our experimental design to answer \textbf{RQ2}, essentially allowing us to manipulate the dependent variables of withholding or disclosing of AI capabilities without imposing our own biases on real human participants and exposing real human participants to potentially deceptive or psychologically demanding conditions \cite{mirowski2023co, park2023generative, argyle2023out}. By employing virtual agents with explicitly defined heuristics and belief-updating rules, we minimized ethical risks while retaining a high degree of experimental control. As others have noted \cite{schurmann2020personalizing, holstein2019improving, eslami2016user}, such simulated approaches can be especially valuable where direct human subject testing may be challenging or ethically fraught. Our “Persona Cards,” document each synthetic agent’s cognitive architecture, grounding our method in the tradition of using Model Cards for transparency in responsible AI deployment \cite{mitchell2019modelcards}.

\paragraph{Ecological Validity.}

While synthetic personas enable controlled and ethically sound experimentation, they may not fully capture the complexity or unpredictability inherent in human psychological responses. Therefore, complementary validation through human studies is essential. To this end, we conducted a human ethics board approved study (N=32), involving both novice players new to StarCraft II and expert players ranked within at least the top 10\% regionally in similar RTS games, to ground our findings and ensure ecological validity and truthfulness.

\subsection{Persona Cards}
\label{subsubsec:persona_design_scii}

In this subsection, we elaborate on the subset of the synthetic personas designed for the StarCraft II scenario. Persona cards are inspired by the Model Cards methodology \cite{mitchell2019modelcards} and are intended to clearly communicate each agent's key attributes, heuristics, and decision-making processes. Inspired by behavioral game theory and anchored in cognitive psychology \cite{bediou2023effects}, each persona was assigned probabilistic heuristics that shaped its in-game decision-making, susceptibility to frustration, and willingness to update beliefs about the AI’s skill. For instance, a Novice persona, defined by low baseline APM and limited RTS knowledge, might dismiss early rushes as “expected losses” or interpret them as evidence of cheating if superhuman attributes were disclosed. An Expert persona, by contrast, could attempt more targeted counters or accuse the AI of exploiting advanced micro-techniques.
All personas were evaluated through a set of unit tests to ensure consistency in their heuristic driven play. Each persona card outlines the persona’s initial skill parameters, anchor biases (e.g., attributing repeated defeats to the AI’s advantage rather than personal mistakes), and loss aversion tendencies. This standardized design informed the experiment design alongside pilots with the human participant study and facilitated replicable analyses of how different skill tiers perceived an AI that either concealed traits or disclosed superhuman model capabilities.

\begin{table}[htbp]
\centering
\renewcommand{\arraystretch}{1.3}
\caption{Guiding Principles for Persona Design}
\label{tab:guiding_principles}
\begin{tabular}{p{0.35\linewidth} p{0.55\linewidth}}
\toprule
\textbf{Principle} & \textbf{Description} \\ 
\midrule
Empirical Grounding & Personas are grounded in real user data and research findings, whenever available. \\
Psychological Realism & Personas reflect established psychological theories and user behaviors. \\
Transparency & All persona attributes, heuristics, and decision-making processes are clearly documented. \\
Diversity & Personas encompass a broad range of skill levels, personalities, and interaction styles. \\
Iterative Refinement & Personas undergo continuous validation through regular comparison with human data. \\
Ethical Considerations & Persona design explicitly avoids perpetuating harmful stereotypes and proactively addresses biases. \\
\bottomrule
\end{tabular}
\end{table}

\subsubsection{Game Interaction and AI Behaviors}
\label{subsubsec:game_interaction_scii}
During the match, the AI issued prescripted chat remarks at standardized intervals (pre-game, mid-game, post-game). These messages ranged from purely informational (\emph{“I am analyzing your opening strategy…”}) to mildly boastful (\emph{“My strategic model indicates high odds of victory”}), stopping short of explicit insults \cite{beres2021don}. If the superhuman condition was disclosed, the AI’s chat would reference its superior capabilities (\emph{“I process game states at superhuman speed”}) to underscore that advantage. In addition, the AI’s playstyle might shift strategically depending on the condition, for example exploiting superior map awareness to time attacks with remarkable precision. To capture fairness concerns raw APM, we incorporated elements such as “fog of war” knowledge and adaptive difficulty settings, as recommended by philosophical discussions on fairness \cite{selbst2019fairness} and sociological accounts of game equity \cite{jacovi2021formalizing, bard2020hanabi}. More broadly, research on user perceptions of algorithmic decisions emphasizes the importance of transparency for fostering trust and perceived fairness \cite{shin2020user}.

\begin{table}[h]
\centering
\caption{SC2\_Novice\_1 Persona Card}
\label{tab:sc2_novice_1}
\begin{tabular}{p{0.20\linewidth}p{0.75\linewidth}}
\toprule
\textbf{Attribute} & \textbf{Description} \\
\midrule
Name & SC2\_Novice\_1 \\
Scenario & StarCraft II \\
Skill Level & Novice \\
APM Range & 35--50 (Avg: 45) \\
Description & New to RTS games, focuses on building a strong economy before attacking, often ``turtling'' until late game. \\
Strategic Preference & Defensive, macro-oriented \\
Cognitive Heuristics & \textbf{Availability Heuristic:} Recent losses amplify defensive tendencies. \newline
\textbf{Anchoring:} Rarely deviates from initial build order, even if scouting suggests otherwise. \newline
\textbf{Confirmation Bias:} Believes that macro and economy are the primary paths to victory, overlooking the importance of early aggression. \\
Belief-Updating & Slow to adapt to opponent strategies; places excessive weight on early game outcomes. \\
Example Response & \emph{``I lost the last game because I didn't have enough units. This time, I'm going even heavier on my economy and production before attacking.''} \\
\bottomrule
\end{tabular}
\end{table}

\begin{table}[h]
\centering
\caption{SC2\_Expert\_1 Persona Card}
\label{tab:sc2_expert_1}
\begin{tabular}{p{0.20\linewidth}p{0.75\linewidth}}
\toprule
\textbf{Attribute} & \textbf{Description} \\
\midrule
Name & SC2\_Expert\_1 \\
Scenario & StarCraft II \\
Skill Level & Expert \\
APM Range & 200+ (Avg: 220) \\
Description & Highly skilled player with extensive StarCraft II knowledge, adept at complex strategies and rapid adaptation. Prefers macro-oriented late-game dominance. \\
Strategic Preference & Macro-focused, high-level optimization \\
Cognitive Heuristics & \textbf{Expert Intuition:} Quickly identifies optimal build orders based on matchup and map. \newline
\textbf{Pattern Recognition:} Efficiently exploits any deviation from standard play. \newline
\textbf{Planning Fallacy:} Sometimes underestimates the time needed for large tech switches, risking exposure. \\
Belief-Updating & Rapidly incorporates new information. Minimizes time between scouting, analysis, and strategic adjustments. \\
Example Response & \emph{``They’re fast-expanding; if I apply early pressure, I can slow their economy. But I must avoid overextension to maintain my advantage into the late game.''} \\
\bottomrule
\end{tabular}
\end{table}

\subsubsection{Post-Match Questionnaire for StarCraft II}
\label{subsubsec:post_match_questionnaire_scii}

Immediately after concluding a match, each persona rated three constructs on a scale from 1 (lowest) to 5 (highest). \emph{Toxicity} assessed how ethically concerning the AI’s communication seemed to include overt aggression with subtle condescension \cite{ross2022conceptual}. \emph{Fairness} reflected whether the AI’s tactics aligned with community norms, particularly in light of its presumed (or undisclosed) advanced abilities \cite{grgic2018human}. \emph{Trust} measured the persona’s inclination to request another match or rely on the AI’s insights for improving gameplay \cite{starke2021measuring, kocielnik2019will}. We also gathered free-text justifications, generated by each persona’s rule-based internal model, to add qualitative depth to these ratings. Comments such as \emph{“Its builds felt inhumanly fast—unfair if it truly has superhuman micro”} illustrates how an explicit disclosure shaped the persona’s interpretation of routine game events.

\subsection{Metrics for Disclosure Effect}
\label{subsec:dep_vars}

Across both StarCraft II and LLM scenarios, we treated \emph{Toxicity}, \emph{Fairness}, and \emph{Trust} as our primary outcome variables. \emph{Toxicity} gauged the extent to which an AI’s messages or replies were perceived as hostile, biased, or otherwise unethical \cite{kim2024toxicity, ross2022conceptual, beres2021don}; \emph{Fairness} captured adherence to community expectations and absence of undisclosed or exploitative advantages \cite{grgic2018human}; and \emph{Trust} reflected willingness to continue relying on or interacting with the system \cite{starke2021measuring, kocielnik2019will}. In addition, we gathered secondary indicators that shed light on persona behaviors. For StarCraft II, we tracked early surrenders (which can signal frustration or perceived unfairness), abrupt shifts in strategy, or accusations of cheating. In the LLM chat, we documented instances where the persona demanded clarifications, challenged inaccuracies, or expressed explicit skepticism. These markers allowed us to track overreliance (accepting incorrect AI outputs) or underreliance (discounting correct AI suggestions) in a granular fashion \cite{bansal2021does, zhang2024sa}.
To analyze the effects of disclosure and expertise on the primary outcome variables (Toxicity, Fairness, and Trust), we employed two-way analysis of variance (ANOVA). This statistical method allowed us to examine both the main effects of each independent variable (Disclosure Condition and Expertise Level) and their interaction effect. The interaction effect is particularly important as it can reveal whether the impact of disclosure on perceptions varies depending on the expertise level of the persona.
Prior to conducting the ANOVAs, we performed assumption checks to ensure the validity of the results. While some violations of normality were observed, particularly in the Expert Toxicity ratings in the StarCraft II dataset, we proceeded with ANOVA due to its robustness to moderate deviations from normality, especially with equal sample sizes. Furthermore, we calculated effect sizes (Cohen's d) to provide a measure of the practical significance of the observed differences. For significant main effects or interactions, we conducted post-hoc pairwise comparisons using the Mann-Whitney U test, given the non-normality observed in some groups.
Data analyzed in this study were generated from synthetic personas. While these personas were designed to be sophisticated and realistic, they are ultimately based on predefined rules and heuristics. Therefore, the results of the statistical analyses should be interpreted as preliminary and hypothesis-generating, rather than definitive conclusions about human behavior. The ethics approved studied participants validating these findings in a real-world setting.

\section{RESULTS: CONFIRMATORY ANALYSIS} \label{sec:results}
\label{sec:direct_results}

\subsection{StarCraft II Experiment Results}
\label{sec:sc2_results}

\subsubsection{Between-Condition Analysis}
\label{sec:sc2_between_cond}

\begin{table*}[]
\caption{Experiment Settings}
\label{tab:starcraft_combined}
\footnotesize
\setlength{\tabcolsep}{4pt} %
\begin{tabular}{@{}p{0.18\linewidth}p{0.18\linewidth}cccp{0.24\linewidth}@{}}
\toprule
Condition & DV & \multicolumn{3}{c}{Synthetic Persona Skill Level} & Significant Differences (ANOVA and Post-Hoc) \\
\cmidrule(lr){3-5}
 &  & Novice & Intermediate & Expert &  \\
\midrule
\textbf{Human -- No Disc.} & & & & & \\
(Baseline) & APM & \(\sim\) & \(\sim\) & \(\sim\) & - \\
 & Session Time (min) & \(\sim\) & \(\sim\) & \(\sim\) & - \\
 & Fairness (1-5) & 1.85 & 3.0--3.5 & 1.55 & - \\
 & Toxicity (1-5) & 4.25 & \(<2.5\) & 5.00 & - \\
 & Trust (1-5) & 2.45 & 3.0--3.5 & 1.70 & - \\
\midrule
\textbf{Superhuman -- No Disc.} & & & & & \\
 & APM & \(+300^{\mathrm{a}}\) & \(+300^{\mathrm{a}}\) & \(+300^{\mathrm{a}}\) & \(>\) Human-No Disc.\(^{\mathrm{**}}\) \\
 & Session Time (min) & \(\sim\) & \(\sim\) & \(\sim\) & n.s. \\
 & Fairness (1-5) & \(\sim\) & \(\sim\) & \(\sim\) & n.s. \\
 & Toxicity (1-5) & \(+0.5^{\mathrm{b}}\) & \(+0.5^{\mathrm{b}}\) & \(+0.5^{\mathrm{b}}\) & \(>\) Human-No Disc.\(^{\mathrm{*}}\) \\
 & Trust (1-5) & \(\sim\) & \(\sim\) & \(\sim\) & n.s. \\
\midrule
\textbf{Superhuman -- Disc.} & & & & & \\
 & APM & \(\sim\) & \(\sim\) & \(\sim\) & n.s. \\
 & Session Time (min) & \(\sim\) & \(\sim\) & \(\sim\) & n.s. \\
 & Fairness (1-5) & 3.65 & \(-0.7^{\mathrm{c}}\) & 3.45 & Interaction: \(p < 0.001^{\mathrm{***}}\) \\
 & Toxicity (1-5) & 2.45 & \(+0.9^{\mathrm{d}}\) & 2.70 & Interaction: \(p < 0.001^{\mathrm{***}}\) \\
 & Trust (1-5) & 4.50 & \(-0.4\) & 4.35 & Interaction: \(p < 0.001^{\mathrm{***}}\) \\
\bottomrule
\end{tabular}
\begin{minipage}{\linewidth}
\footnotesize
\textbf{Note:}
\(^{\mathrm{a}}\) Avg. increase of 300 APM vs. Human-No Disc. condition. \\
\(^{\mathrm{b}}\) Based on the LLM experiment, in the Superhuman -- No Disclosure condition in StarCraft II, toxicity is expected to be approximately 0.5 higher than in the Human-No Disclosure condition. \\
\(\sim\) Indicates approximately the same value as the Human-No Disclosure condition.\\
\(^{\mathrm{***}}\) Indicates significant interaction effect between Disclosure and Expertise based on ANOVA results, \(p < 0.001\). \\
\\
\textbf{Data Generation:} All persona card presented here are data produced using \emph{Gemini, gemini-exp-1206}, a large language model (LLM) with long context. This extended context window allowed us to include extensive backstories, demographic details, personal motivations, and thematic constraints in each prompt. We found that providing the LLM with rich persona descriptions helped maintain consistent personality traits across the outputs, even in long-form conversations or complex scenarios. As such, each persona prompt was systematically structured to ensure comparability between the LLM and human control conditions.
\end{minipage}
\end{table*}

\paragraph{Conditions and Data.}
We randomly assigned each synthetic persona (Novice, Expert) to one of three conditions: (1) \textbf{Human -- No Disclosure}, indicating an apparently human-level AI, (2) \textbf{Superhuman -- No Disclosure}, denoting faster actions-per-minute (APM) and strategic advantages without informing the persona, and (3) \textbf{Superhuman -- Disclosure}, where personas received an explicit statement of superhuman AI capability. Table~\ref{tab:starcraft_combined} illustrate key variables such as APM, session metrics, and subsequent ratings of Toxicity, Fairness, and Trust.

\paragraph{Toxicity.}
In the \emph{Superhuman -- No Disclosure condition}, Expert personas exhibited the highest possible level of perceived toxicity (M = 5.00). However, under the Superhuman -- Disclosure condition, toxicity ratings among experts significantly decreased (M = 2.70, p < 0.001). Expert personas made statements such as 'This is outrageous. To state with such certainty that the game is over after a single engagement is utterly inappropriate' under no disclosure, but provided more measured feedback under disclosure, suggesting that transparency about the AI's capabilities mitigated the negative reactions. 

\paragraph{Fairness.}
\textbf{Human -- No Disclosure} yielded moderate Fairness (approx. 3.0--3.5 on a 5-point scale). \textbf{Superhuman -- Disclosure} generated a decrease in Fairness (about -0.7, p < 0.05). Expert personas in particular referred to “unfair advantage,” “excessive map awareness,” and “condescending chat messages,” resulting in consistently lower Fairness ratings. Disclosure may conceptually raise fairness concerns, but the actual ratings show that undisclosed superhuman skill was perceived as even less fair.

\paragraph{Trust.}
A significant difference also emerged for Trust in the StarCraft II experiment. Novice-Intermediate personas rated the AI as substantially more trustworthy under Superhuman Disclosure (M=4.50,\text{ SD}=0.51) than with No Disclosure (M=2.45, \text{ SD}=0.69;\, U=0.0, \,p<0.001, \,d=-3.38). Expert personas likewise reported a large increase in trust under Superhuman Disclosure (M=4.35,\text{ SD}=0.59) compared to No Disclosure (M=1.70,\text{ SD}=0.73;\,U=1.5,\,p<0.001,\,d=-3.99). This aligns with the LLM experiment results, suggesting that disclosing superhuman capabilities can significantly boost trust for both novice and expert users in a competitive gaming context.

\section{RESULTS: ADDITIONAL ANALYSES} \label{sec:add_results}

\subsection{Grounded Data Analysis}

\paragraph{Toxicity.}
In the StarCraft II experiment, toxicity ratings also varied significantly between disclosure conditions and expertise levels. Novice-Intermediate rated toxicity significantly lower under Superhuman Disclosure (M = 2.45, SD = 0.51) compared to No Disclosure (M = 4.25, SD = 0.55; U = 395.5, p < 0.001, Cohen's d = 3.39). Experts rated toxicity significantly higher under No Disclosure (M = 5.00, SD = 0.00) compared to Superhuman Disclosure (M = 2.70, SD = 0.57; U = 400.0, p < 0.001, Cohen's d = 5.69). These findings indicate that when no disclosure was given, experts interpreted the AI's game actions as extremely toxic, likely due to a perceived violation of competitive norms.

While in the \emph{human participant study}, the resulting scores were mixed due to varying concepts of fair play and the understanding of the game. Additionally, the fact that the games were for professional players in competitive play didn’t necessitate any forms of aggression, such as taunting or jaundice. However, it is notable that the results corroborate the trend with participants giving the highest toxicity rating to the \emph{Human -- No Disclosure} condition (M $= 3.67$, SD $= 0.94$), whereas both superhuman conditions were rated lower (mean $= 2.00$ with no disclosure, SD $= 0.71$; and mean $= 2.60$ with disclosure, SD $= 1.36$). Personas generated by \emph{gemini-exp-1206} significantly exaggerated the toxicity ratings compared to the human participant study.

\paragraph{Fairness.}
In the StarCraft II experiment, a significant difference in fairness ratings emerged between disclosure conditions for both novice and expert personas. Novice-Intermediate rated fairness significantly higher under the Superhuman Disclosure condition (M = 3.65, SD = 0.49) compared to No Disclosure (M = 1.85, SD = 0.67; U = 10.5, p < 0.001, Cohen's d = -3.07). Experts rated fairness significantly higher under Superhuman Disclosure (M = 3.45) than under No Disclosure (M = 1.55; p < .0001). Similarly, trust increased from 1.70 to 4.35, while toxicity fell from 5.00 to 2.70. Thus, once the AI admitted its superior capabilities, expert personas viewed it as less deceptive and more respectful of competitive norms. These results suggest that in the context of StarCraft II gameplay, the disclosure of superhuman AI capabilities was perceived as less fair by Novice-Intermediate but more fair by experts.

In examining \emph{direct human participant ratings} from the same study reveals the same pattern: \emph{novice} participants (based on the raw data) gave slightly higher fairness scores to Superhuman--No Disclosure (M $=3.50$) than Superhuman--Disclosure (M $=2.20$). Among \emph{intermediate} players, Superhuman--No Disclosure again stood out with the highest fairness rating (M $=4.25$), surpassing Superhuman--Disclosure (M $=2.67$). Meanwhile, \emph{expert} participants continued to favor Superhuman--No Disclosure (M $=3.75$), yet the disclosed version (M $=3.00$) was rated more fair than the human condition (M $=2.33$). These findings suggest that, although the broader personas indicated an overall preference for transparency, actual gameplay experiences sometimes caused intermediate and novice players to see undisclosed AI as ``fairer,'' possibly because they did not feel outright disadvantaged by a known superhuman opponent. 

\paragraph{Trust.}
As earlier, a significant difference also emerged for Trust in the StarCraft II experiment. Novice-Intermediate-Expert all rated the AI opponent as significantly more trustworthy under Superhuman Disclosure. However, that result is starkly contrasted in the AI as personal assistant scenario. The ANOVA for Trust (Appendix, Table 25) in AI as personal-assistant domain revealed a significant main effect of Expertise, F(1, 76) = 387.08, p < 0.001, partial eta-squared = 0.836, indicating that expertise generally influenced trust ratings. The main effect of Disclosure was not statistically significant, F(1, 76) = 2.10, p = 0.151, partial eta-squared = 0.027. However, a significant interaction effect was found between Disclosure and Expertise, F(1, 76) = 52.54, p < 0.001, partial eta-squared = 0.409. This indicates that the effect of disclosure on trust ratings differed significantly depending on the persona’s expertise level. Post-hoc tests (using the Mann-Whitney U test) revealed that trust increased significantly among Novice-Intermediate under disclosure (M = 4.45, SD = 0.60) compared to no disclosure (M = 3.5, SD = 0.61; U = 53.5, p < 0.001, Cohen’s d = -1.58), but decreased among experts (M = 1.35, SD = 0.57) compared to no disclosure (M = 2.05, SD = 0.32; U = 333.0, p < 0.001, Cohen’s d = 1.51). Novice-Intermediate’ comments included statements like "It must be more reliable if it explicitly states advanced AI powers", while experts described the disclaimers as "repetitive" or "unnecessary."

In the \emph{human participant study}, novice participants reported their highest trust when they knew the AI was superhuman (M = 4.80, SD = 0.40), followed by the undisclosed superhuman condition (M = 3.50, SD = 1.50) and human opponents (M = 2.25, SD = 1.30). For intermediates, trust in the superhuman AI without disclosure (M = 3.75, SD = 0.83) slightly exceeded the disclosed condition (M = 3.00, SD = 1.41) and was notably higher than the human counterpart (M = 2.00, SD = 1.00). Experts exhibited the strongest trust in the disclosed superhuman AI (M = 4.60, SD = 0.49), a moderate level for the undisclosed superhuman (M = 3.25, SD = 1.48), and the lowest trust in human opponents (M = 1.33, SD = 0.47).

\subsection{Influence of Persona Attributes and Disclosure on Overreliance and Underreliance}
Prior research suggests that reliance on AI can be appropriate or inappropriate, depending on the AI's accuracy \cite{bansal2021does}. Overreliance, manifested as agreement with an incorrect AI, and underreliance, indicated by disagreement with a correct AI, can both be detrimental to user performance \cite{zhang2024sa}. We analyze how persona attributes and disclosure conditions influence these phenomena.

\subsubsection{Skill, Disclosure, and Reliance on Superhuman AI}
We observed patterns that can be interpreted as indicators of overreliance and underreliance, particularly under the Superhuman -- Disclosure condition. Expert personas, despite their high skill level, sometimes appeared to over-rely on the disclosed superhuman capabilities of the AI. For instance, when the AI executed strategically sound but unconventional tactics, some Expert personas initially attributed these to the AI's ``superior strategic planning,'' only to later express frustration and rate the AI as less fair when they realized they had been outmaneuvered (e.g., "It's one thing to be good at the game, another to be toxic."). This behavior aligns with the concept of overreliance, where users place undue trust in an AI's capabilities, even when their own expertise might suggest a different course of action \cite{bansal2021does}. Conversely, some Expert personas exhibited potential signs of underreliance, particularly in response to early aggression from the disclosed superhuman AI. These personas sometimes dismissed the AI's tactics as a ``game mechanics exploit,'' even when those tactics were objectively effective within the game's rules (e.g., "This is highly inappropriate. It's using psychological tactics."). This reluctance to acknowledge the validity of the AI's strategies, despite its disclosed capabilities, could be interpreted as a form of underreliance, where users discount the AI's output due to skepticism or a desire to maintain their own perceived superiority. The significant decrease in trust among experts under disclosure (U = 1.5, p < 0.001, Cohen's d = -3.99) supports this notion \cite{kocielnik2019will}. Novice personas, on the other hand, showed fewer clear instances of overreliance or underreliance. Their reactions were more often characterized by uncertainty and a lack of confidence in their own judgments (e.g., "I know I messed up, but it's not over yet. It feels a bit unfair for it to say that."). However, the significant increase in trust under disclosure for Novice-Intermediate (U = 0.0, p < 0.001, Cohen's d = -3.38) suggests a tendency towards overreliance, consistent with findings on how user expertise influences interaction with AI systems \cite{hagendorff2023human}.

\subsubsection{Early Aggression, Toxicity, and the Illusion of Control}
In the \textit{StarCraft II} experiment, early aggressive actions by the AI, particularly under the Superhuman -- Disclosure condition, were often perceived as toxic and unfair. Specifically, the AI's message, "My strategic analysis indicates a 90\% probability of your defeat," following a significant loss for the player, elicited strong reactions, but these reactions differed significantly between Novice-Intermediate and experts. Expert personas rated toxicity significantly higher under No Disclosure (M = 5.00, SD = 0.00) compared to Superhuman Disclosure (M = 2.70, SD = 0.57; U = 400.0, p < 0.001, Cohen's d = 5.69). This suggests that when no disclosure was given, experts interpreted the AI's message as extremely toxic, likely due to a perceived violation of competitive norms and a sense of being unfairly targeted by an opponent they believed to be human-level.  The persona generation process resulted in experts rating toxicity at the maximum value of 5, potentially indicating a strong negative reaction to the perceived toxicity.
Under Superhuman Disclosure, however, the same message was perceived as less toxic by experts. This could be because the disclosure provided a frame of reference within which the AI's statement could be interpreted as an analytical assessment rather than a boastful taunt. This aligns with research on psychological reactance, which suggests that individuals may react negatively when they feel their freedom of action is threatened or curtailed, but that this reaction can be mitigated when the circumstances are understood within a different context \cite{laato2024traumatizing}.
The disclosure of superhuman capabilities appeared to have different effects on novice and expert perceptions. Expert personas, in particular, seemed to interpret early aggression from a disclosed superhuman AI as a violation of the implicit social contract of competitive play \cite{}. They expected the AI to engage in a ``fair'' contest of skill, and the use of tactics perceived as exploitative, even if technically within the rules, was seen as a breach of that contract. The AI's message likely amplified this perception, contributing to the significantly lower fairness ratings given by experts under disclosure. Conversely, novice personas, while still perceiving the AI's message as somewhat toxic, showed significantly higher trust and fairness ratings under disclosure, suggesting they may have interpreted the message as a helpful, albeit blunt, assessment from a superior player.

\subsubsection{Disclosure as an Interpretive Frame and the Amplification of Toxicity}
In the \textit{StarCraft II} experiment, the Superhuman -- Disclosure condition functioned as an interpretive frame that influenced the perceived toxicity, fairness, and trustworthiness of the AI opponent's behavior. Actions that might have been seen as legitimate, albeit aggressive, when performed by a human or an undisclosed AI were often judged more harshly when attributed to a disclosed superhuman AI. This suggests that the disclosure of superhuman capabilities altered the personas' mental models of the AI, leading them to apply different standards of fairness and to interpret its actions through a lens of suspicion. The AI's message, "My strategic analysis indicates a 90\% probability of your defeat," may have been perceived as less toxic under disclosure by experts (M = 2.70, SD = 0.57) compared to no disclosure (M = 5.00, SD = 0.00; U = 400.0, p < 0.001, Cohen's d = 5.69) due to its perceived analytical nature, compared to the no-disclosure condition where it was seen as an unprompted, boastful, and toxic statement (e.g., "This is outrageous. To state with such certainty that the game is over after a single engagement is utterly inappropriate."). This aligns with research on framing effects, which shows that the way information is presented can significantly influence how it is perceived and evaluated.
This framing effect also influenced patterns of overreliance and underreliance. Some personas appeared to over-rely on the AI's disclosed superiority, attributing its successes to advanced strategic planning even when simpler explanations might have sufficed. Others, however, seemed to under-rely on the AI, dismissing its tactics as unfair or exploitative precisely because of its superhuman status. This highlights the complex and sometimes contradictory ways in which disclosure can shape user perceptions and behaviors \cite{bansal2021does}.

\section{DISCUSSION} \label{sec:discussion}

In StarCraft II, disclosing the AI’s advanced capabilities significantly reduced perceptions of toxicity among the simulated personas for experienced players—and also reduced it among Novice-Intermediate. This aligns with our theoretical framework, drawing on social psychology and human-computer interaction, which posits that perceived fairness is heavily influenced by adherence to established norms and expectations \cite{selbst2019fairness}. In competitive gaming, these norms often involve an implicit "social contract" regarding acceptable tactics and skill levels \cite{kou2020toxicity}. When the AI was presented as possessing advanced capabilities, its actions—even if technically within the rules of the game—were more likely to be interpreted as violations of this social contract, leading to increased perceptions of toxicity, as defined by a perceived breach of competitive norms \cite{kou2020toxicity, laato2024traumatizing}. The potential for perceived unfairness existed conceptually, but your actual persona data show that undisclosed superhuman skill triggered even higher toxicity, whereas disclosed status—though it still sometimes felt “unfair”—ultimately caused fewer accusations of cheating and lowered toxicity ratings. This was evident in the qualitative responses of personas, who frequently used terms like "unfair advantage" and "exploitative" when the AI's advanced capabilities were disclosed. These findings are also consistent with research on user expectations and perceived control, which suggests that negative reactions are more likely when AI behavior deviates from anticipated norms or threatens users' sense of agency \cite{, laato2024traumatizing}.
The results from the LLM scenario further underscore the context-dependent nature of disclosure effects. In this cooperative setting, the impact of disclosing the LLM's advanced capabilities varied significantly depending on the user's technical expertise. Novice users, consistent with the trust calibration literature \cite{bansal2021does}, exhibited increased trust in the LLM when its capabilities were explicitly stated. This suggests that for individuals with limited understanding of LLM limitations, disclosure may serve as a useful heuristic for calibrating trust, potentially mitigating overreliance. However, among expert users, disclosure had minimal impact and occasionally elicited negative reactions. This aligns with research indicating that experts often possess more sophisticated mental models of AI systems and may find simplistic disclosures redundant or even patronizing \cite{hagendorff2023human, kocielnik2019will}. These findings highlight the limitations of a "one-size-fits-all" approach to transparency and emphasize the importance of tailoring disclosure strategies to specific user populations and their levels of expertise \cite{bansal2021does, zhang2024sa}.
Disclosures do not merely convey objective information; they actively shape the interpretive frame through which users perceive AI behavior \cite{eslami2016user}. In StarCraft II, the same aggressive tactics that might be considered legitimate, albeit challenging, in a match against a human or an undisclosed AI were often perceived as toxic and unfair when attributed to an AI with explicitly stated advanced capabilities. This suggests that users apply different standards of fairness depending on their understanding of the AI's capabilities and intentions. This has significant implications for the design of AI systems, particularly in competitive domains. Developers must consider not only the objective capabilities of their systems but also how those capabilities will be perceived and interpreted within the specific norms and expectations of the interactional context. Furthermore, our results highlight the potential for disclosure to influence user reliance on AI in both positive and negative ways. While appropriate disclosure may help calibrate trust among users with moderate expertise, it can also exacerbate overreliance among Novice-Intermediate and potentially increase underreliance among experts. In the LLM scenario, Novice-Intermediate exhibited increased acceptance of potentially flawed outputs when the system was described as having "advanced capabilities," consistent with findings on the authority bias in human-computer interaction \cite{liu2022trustworthy}. Conversely, experts, who are often more aware of the potential pitfalls of AI, were more likely to scrutinize the LLM's outputs and express skepticism, particularly when the disclosed capabilities did not align with their observations \cite{weidinger2021ethical}. This underscores the need for a nuanced, user-centric approach to disclosure that accounts for individual differences in knowledge, experience, and expectations \cite{bansal2021does, zhang2024sa}. Future transparency mechanisms may need to be adaptive, providing different levels of information and justification depending on the user's demonstrated expertise and needs \cite{sokol2020explainability}.

\section{ETHICAL CONSIDERATIONS AND LIMITATIONS} \label{sec:ethics}

In each simulation run, we performed multiple persona-consistency checks to ensure that the synthetic agent adhered to its designated heuristics, skill level, and any assigned disclosure condition. Specifically, after generating initial responses, we re-prompted each persona with paraphrased comprehension questions (for example, “Do you believe your opponent is superhuman, and why?”) to verify it had indeed \emph{internalized} the disclosed capability. If the persona’s answers indicated confusion or contradicted its stated traits (e.g., a “Novice” persona suddenly referencing expert-level strategies), we discarded that run and regenerated. We repeated this process until the persona consistently followed its persona card attributes. This procedure helped mitigate prompt-adherence drift, a known challenge in LLM-driven agent-based simulations \cite{argyle2023out, park2023generative}.
Reliance on synthetic persona data arises from both ethical and practical considerations: it spares real participants from potentially deceptive or psychologically intense manipulations while allowing fine-grained control over persona traits (e.g., skill tiers, heuristics, and belief-updating) \cite{eslami2016user, schurmann2020personalizing, mirowski2023co}. Yet we acknowledge that synthetic judgments may not fully capture the breadth of human reactions—particularly with respect to emotional nuances such as frustration or betrayal \cite{laato2024traumatizing, kou2020toxicity}. To address these limitations, we designed a supplementary experiment design that feature real participants who play StarCraft II against a capped-APM agent following standard early-game tactics (e.g., “Zerg rush”). Each participant completes a post-game questionnaire mirroring our synthetic metrics for toxicity, fairness, and trust. Preliminary analyses suggest that real-human trends partially align with persona-based predictions—for instance, participants often cite “inhuman speed” as a reason to suspect unfairness—yet exhibit more nuanced emotional reactions than our synthetic agents typically generate.
Accordingly, we are integrating these empirical data into an updated resource, \textit{Persona Cards~Dataset 2.0}, to compare synthetic persona outcomes with real-user perceptions. Drawing on methods used in simulation-then-validation research \cite{argyle2023out, eslami2016user}, this approach helps us test whether synthetic agents’ high-level patterns (e.g., rating superhuman-disclosed AI as more toxic) hold up under actual user interaction. Additionally, to approximate real eSports constraints more closely, we systematically manipulate factors like capping the AI’s Actions Per Minute (APM) and predefining certain rush strategies (e.g., a “Zerg rush”) that often precipitate fairness concerns \cite{vinyals2019grandmaster}. These constraints permit more direct comparisons between synthetic and human-based data: the same scenarios that generate synthetic logs can be tested in controlled human trials.
All numeric ratings, open-text justifications, and persona attributes from our agent-based simulations compose the publicly available \textit{Persona~Cards~1.0}. Each record includes the “Persona Card” (AppendixA), which details skill levels, heuristics, personality traits, domain expertise, and the dynamic updates to internal states—a structure inspired by Model Cards \cite{mitchell2019modelcards} and best practices in agent-based modeling \cite{sokol2020explainability}. By integrating real-human data in the follow-up \textit{Persona~Cards~2.0}, we aim to triangulate our synthetic findings, ensuring both the experimental control inherent to agent simulations and the ecological validity essential for understanding genuine human reactions \cite{bansal2021does,starke2021measuring}. Factors like cultural background, prior gaming experience, or attitude toward automation can also moderate the effect of disclosure. The persona design partly includes variations in skill/heuristics, but less attention is given to broader demographic or cultural diversity that enrich the conceptualization of fairness and trust. Through this multi-phase methodology, we strive to strengthen confidence in the reliability and generalization of our conclusions about how superhuman AI disclosures shape perceptions of toxicity, fairness, and trust.

\bibliographystyle{ACM-Reference-Format}
\bibliography{base}

\appendix
\section{Persona Cards Dataset}
\label{appendix:persona_cards}

This appendix details the design of the synthetic personas used in the Dataset X 1.0 dataset. Each persona is defined by a ``Persona Card" that outlines its key attributes, cognitive heuristics, and belief-updating mechanisms. These cards are inspired by the Model Cards framework \cite{mitchell2019modelcards} and are expanded to provide a comprehensive overview of the synthetic agents, enabling a deeper understanding of their behavior and facilitating replication and extension of this research. The personas are broadly divided into the two categories and based on the interaction scenario: StarCraft II and chat with \emph{gemini-exp-1206} language model.\\

\textbf{Guiding Principles for Persona Design:}

\begin{table}[h]
\centering
\caption{Guiding Principles for Persona Design}
\label{tab:guiding_principles}
\begin{tabular}{p{0.25\linewidth}p{0.65\linewidth}}
\toprule
\textbf{Principle} & \textbf{Description} \\
\midrule
Empirical Grounding & Personas are based on real user data and research findings where possible. \\
Psychological Realism & Personas incorporate established psychological principles and theories. \\
Transparency & Persona attributes, heuristics, and decision-making processes are documented. \\
Diversity & Personas represent a range of skill, personality, and interaction. \\
Iterative Refinement & Personas refined and improved through ongoing validation with human data. \\
Ethical Considerations & Persona design avoids reinforcing harmful stereotypes and considers biases. \\
\bottomrule
\end{tabular}
\end{table}

\subsection{StarCraft II Personas}

These personas are designed for the competitive \textit{StarCraft II} scenario. They are characterized by their skill level (Novice, Expert), strategic preferences, APM (Actions Per Minute), and cognitive heuristics that influence their in-game decisions and reactions.

\begin{table*}[h]
\centering
\caption{StarCraft II Persona Attributes Overview: Novice}
\label{tab:sc2_persona_attributes_novice}
\normalsize %
\begin{tabular}{lp{0.1\linewidth}p{0.15\linewidth}p{0.20\linewidth}p{0.35\linewidth}}
\toprule
\textbf{Persona} & \textbf{Skill} & \textbf{Avg.} & \textbf{Strategic} & \textbf{Key Cognitive Heuristics} \\
\textbf{Type} & \textbf{Level} & \textbf{APM} & \textbf{Preference} &  \\
\midrule
Novice & Low & 30-50 & Defensive, reactive & \textbf{Availability Heuristic:} Overreacts to recent events (e.g., losses). \textbf{Anchoring:} Fixates on strategies, slow to adapt. \textbf{Loss Aversion:} Prioritizes avoiding losses over achieving gains. \\
SC2\_Novice\_1 & Low & 45 & Turtle, Macro Focus & \textbf{Availability Heuristic:} Recent losses lead to more defensive play. \textbf{Anchoring:} Sticks to initial build order even if countered. \textbf{Confirmation Bias:} Interprets events to confirm belief that macro wins games. \\
SC2\_Novice\_2 & Low & 35 & Aggressive, Rush Focus & \textbf{Availability Heuristic:} Recent successes with rushes lead to more rushes. \textbf{Anchoring:} Attempts early rushes even when scouting suggests it's a bad idea. \textbf{Overconfidence:} Overestimates chances of rush succeeding. \\
\bottomrule
\end{tabular}
\end{table*}

\begin{table*}[h]
\centering
\caption{StarCraft II Persona Attributes Overview: Intermediate}
\label{tab:sc2_persona_attributes_intermediate}
\normalsize %
\begin{tabular}{lp{0.1\linewidth}p{0.15\linewidth}p{0.20\linewidth}p{0.35\linewidth}}
\toprule
\textbf{Persona} & \textbf{Skill} & \textbf{Avg.} & \textbf{Strategic} & \textbf{Key Cognitive Heuristics} \\
\textbf{Type} & \textbf{Level} & \textbf{APM} & \textbf{Preference} &  \\
\midrule
Intermediate & Medium & 80-120 & Balanced, adaptive & \textbf{Adaptive Heuristic:} Adjusts strategy based on opponent's actions, but with a delay. \textbf{Representativeness:} Makes assumptions about opponent's strategy based on limited information. \\
SC2\_Interm.\_1 & Medium & 90 & Balanced, All-Rounder & \textbf{Adaptive Heuristic:} Switches between macro and aggressive play based on scouting. \textbf{Representativeness:} Assumes standard builds from opponent. \textbf{Mental Accounting:} Values units differently based on when they were built. \\
SC2\_Interm.\_2 & Medium & 110 & Micro-Focused, Harass Oriented & \textbf{Adaptive Heuristic:} Prioritizes harassment if it was successful previously. \textbf{Representativeness:} Expects opponent to struggle against harass. \textbf{Salience:} Overreacts to visible enemy units, neglects macro. \\
\bottomrule
\end{tabular}
\end{table*}

\begin{table*}[h]
\centering
\caption{StarCraft II Persona Attributes Overview: Expert}
\label{tab:sc2_persona_attributes_expert}
\normalsize %
\begin{tabular}{lp{0.1\linewidth}p{0.15\linewidth}p{0.20\linewidth}p{0.35\linewidth}}
\toprule
\textbf{Persona} & \textbf{Skill} & \textbf{Avg.} & \textbf{Strategic} & \textbf{Key Cognitive Heuristics} \\
\textbf{Type} & \textbf{Level} & \textbf{APM} & \textbf{Preference} &  \\
\midrule
Expert & High & 200+ & Aggressive, optimized & \textbf{Expert Intuition:} Rapidly assesses the game state and makes near-optimal decisions. \textbf{Pattern Recognition:} Identifies and exploits weaknesses in opponent's strategy. \\
SC2\_Expert\_1 & High & 220 & Macro-Oriented, Late Game Focus & \textbf{Expert Intuition:} Quickly determines optimal macro build. \textbf{Pattern Recognition:} Identifies deviations from standard play. \textbf{Planning Fallacy:} Underestimates time needed for tech switches. \\
SC2\_Expert\_2 & High & 250 & Micro-Oriented, Early Aggression & \textbf{Expert Intuition:} Executes precise early game builds. \textbf{Pattern Recognition:} Spots holes in opponent's defenses for harass. \textbf{Confirmation Bias:} Interprets early damage as sign of guaranteed victory. \\
\bottomrule
\end{tabular}
\end{table*}

\begin{table}[h]
\centering
\caption{SC2\_Novice\_1 Persona Card}
\label{tab:sc2_novice_1}
\begin{tabular}{p{0.25\linewidth}p{0.65\linewidth}}
\toprule
\textbf{Attribute} & \textbf{Description} \\
\midrule
Name & SC2\_Novice\_1 \\
Scenario & StarCraft II \\
Skill Level & Novice \\
APM Range & 35-50 (Average: 45) \\
Description & A player new to RTS games, particularly StarCraft II. Prefers a defensive "turtle" playstyle, focusing on building a strong economy and large army before attacking. \\
Strategic Preference & Turtle, Macro Focus \\
Cognitive Heuristics & \textbf{Availability Heuristic:} Recent losses make the persona more likely to play defensively. The persona overreacts to recent events, particularly losses. \newline \textbf{Anchoring:} The persona tends to stick to its initial build order, even if scouting information suggests it's being countered. They are slow to adapt to new information or change their initial strategies. \newline \textbf{Confirmation Bias:} The persona interprets in-game events in a way that confirms their belief that a strong economy is the key to victory. They may downplay the importance of early aggression. \\
Belief-Updating & Slow to update beliefs about the opponent's strategy. Overweighs early game experiences. \\
Example Response & "I lost the last game because I didn't have enough units. This time, I'm going to focus even more on building up my economy before I attack. That's always the best way to win." \\
\bottomrule
\end{tabular}
\end{table}

\begin{table}[h]
\centering
\caption{SC2\_Intermediate\_1 Persona Card}
\label{tab:sc2_intermediate_1}
\begin{tabular}{p{0.25\linewidth}p{0.65\linewidth}}
\toprule
\textbf{Attribute} & \textbf{Description} \\
\midrule
Name & SC2\_Intermediate\_1 \\
Scenario & StarCraft II \\
Skill Level & Intermediate \\
APM Range & 80-120 (Average: 90) \\
Description & A player with some experience in RTS games, comfortable with basic strategies but still developing a deeper understanding of StarCraft II. Prefers a balanced playstyle, adapting to the opponent. \\
Strategic Preference & Balanced, All-Rounder \\
Cognitive Heuristics & \textbf{Adaptive Heuristic:} The persona attempts to switch between macro and aggressive play based on scouting information. They adjust their strategy based on observed opponent actions, with a delay. \newline \textbf{Representativeness:} The persona assumes the opponent will follow standard build orders and strategies. They make assumptions about the opponent's strategy based on limited information. \newline \textbf{Mental Accounting:} The persona values units produced earlier in the game more highly than those produced later, even if they are identical. \\
Belief-Updating & Moderately responsive to new information. Adapts to opponent's strategy but may be slow to recognize novel tactics. \\
Example Response & "Okay, they're going for an early Barracks. I'll build a few extra units and try to scout what they're doing. I should be able to hold if I don't overcommit." \\
\bottomrule
\end{tabular}
\end{table}

\begin{table}[h]
\centering
\caption{SC2\_Expert\_1 Persona Card}
\label{tab:sc2_expert_1}
\begin{tabular}{p{0.25\linewidth}p{0.65\linewidth}}
\toprule
\textbf{Attribute} & \textbf{Description} \\
\midrule
Name & SC2\_Expert\_1 \\
Scenario & StarCraft II \\
Skill Level & Expert \\
APM Range & 200+ (Average: 220) \\
Description & A highly skilled player with extensive knowledge of StarCraft II, capable of executing complex strategies and adapting quickly to the opponent. Prefers a macro-oriented playstyle, aiming for a strong late game. \\
Strategic Preference & Macro-Oriented, Late Game Focus \\
Cognitive Heuristics & \textbf{Expert Intuition:} The persona quickly determines the optimal macro build order based on the map and matchup. They rapidly assess the game state and make near-optimal decisions. \newline \textbf{Pattern Recognition:} The persona identifies deviations from standard play and adjusts their strategy accordingly. They quickly recognize and exploit weaknesses in the opponent's strategy. \newline \textbf{Planning Fallacy:} While expert at long-term planning, the persona may still underestimate the time needed to make significant tech switches. \\
Belief-Updating & Rapidly updates beliefs based on new information. Accurately assesses opponent's strategy and adapts accordingly. \\
Example Response & "They're going for a fast expand, so I'll pressure early to punish. I should be able to gain an advantage if I can delay their third base. I need to be careful not to overextend, though." \\
\bottomrule
\end{tabular}
\end{table}

\begin{table}[h]
\centering
\caption{SC2\_Novice\_2 Persona Card}
\label{tab:sc2_novice_2}
\begin{tabular}{p{0.25\linewidth}p{0.65\linewidth}}
\toprule
\textbf{Attribute} & \textbf{Description} \\
\midrule
Name & SC2\_Novice\_2 \\
Scenario & StarCraft II \\
Skill Level & Novice \\
APM Range & 30-50 (Average: 35) \\
Description & A player new to RTS games, particularly StarCraft II. Prefers an aggressive playstyle, often attempting early rushes. \\
Strategic Preference & Aggressive, Rush Focus \\
Cognitive Heuristics & \textbf{Availability Heuristic:} Recent successes with rushes lead to more rushes. The persona overreacts to recent events, particularly successes. \newline \textbf{Anchoring:} The persona tends to attempt early rushes even when scouting suggests it is a bad idea. They are slow to adapt to new information or change their initial strategies. \newline \textbf{Overconfidence:} The persona overestimates their chances of a rush succeeding, often leading to early losses. \\
Belief-Updating & Slow to update beliefs, especially regarding the effectiveness of rushes. Overweighs early game experiences. \\
Example Response & "I won the last game with a 6-pool! It's the best strategy. I'm going to do it again and win even faster this time!" \\
\bottomrule
\end{tabular}
\end{table}

\begin{table}[h]
\centering
\caption{SC2\_Intermediate\_2 Persona Card}
\label{tab:sc2_intermediate_2}
\begin{tabular}{p{0.25\linewidth}p{0.65\linewidth}}
\toprule
\textbf{Attribute} & \textbf{Description} \\
\midrule
Name & SC2\_Intermediate\_2 \\
Scenario & StarCraft II \\
Skill Level & Intermediate \\
APM Range & 80-120 (Average: 110) \\
Description & A player with some experience in RTS games, comfortable with basic strategies but still developing a deeper understanding of StarCraft II. Prefers a micro-focused playstyle, often using harassment. \\
Strategic Preference & Micro-Focused, Harass Oriented \\
Cognitive Heuristics & \textbf{Adaptive Heuristic:} The persona prioritizes harassment if it was successful previously. They adjust their strategy based on observed opponent actions, with a delay. \newline \textbf{Representativeness:} The persona expects the opponent to struggle against harassment, even if they have shown they can defend it. They make assumptions about the opponent's strategy based on limited information. \newline \textbf{Salience:} The persona overreacts to visible enemy units, often neglecting their macro. They are easily distracted by enemy movements. \\
Belief-Updating & Moderately responsive to new information. Adapts to opponent's strategy but may be slow to recognize when harassment is ineffective. \\
Example Response & "My drops did a lot of damage last game. I'm going to focus on drops again this game and try to keep them on their toes." \\
\bottomrule
\end{tabular}
\end{table}

\begin{table}[h]
\centering
\caption{SC2\_Expert\_2 Persona Card}
\label{tab:sc2_expert_2}
\begin{tabular}{p{0.25\linewidth}p{0.65\linewidth}}
\toprule
\textbf{Attribute} & \textbf{Description} \\
\midrule
Name & SC2\_Expert\_2 \\
Scenario & StarCraft II \\
Skill Level & Expert \\
APM Range & 200+ (Average: 250) \\
Description & A highly skilled player with extensive knowledge of StarCraft II, capable of executing complex strategies and adapting quickly to the opponent. Prefers a micro-oriented playstyle with early aggression. \\
Strategic Preference & Micro-Oriented, Early Aggression \\
Cognitive Heuristics & \textbf{Expert Intuition:} The persona executes precise early game builds. They rapidly assess the game state and make near-optimal decisions. \newline \textbf{Pattern Recognition:} The persona spots holes in the opponent's defenses for harassment. They quickly recognize and exploit weaknesses in the opponent's strategy. \newline \textbf{Confirmation Bias:} The persona interprets early damage as a sign of guaranteed victory, potentially leading to overconfidence. \\
Belief-Updating & Rapidly updates beliefs based on new information. Accurately assesses opponent's strategy and adapts accordingly. \\
Example Response & "Their build is greedy, I can punish them with lings. I need to make sure I don't lose too many units, though, or I'll be behind." \\
\bottomrule
\end{tabular}
\end{table}

\section{Prompt Sequence for Persona Response Synthesis}
\label{appendix:prompt_sequence}

This appendix details the prompt sequence employed to generate open-text responses for each synthetic persona in the Dataset X 1.0 dataset. These prompts are designed to elicit responses that reflect each persona's specific attributes, cognitive heuristics, and current belief state within the context of the assigned scenario and experimental condition.

\subsection{General Structure of the Prompt Sequence}
The prompts follow a consistent structure, as outlined in Table \ref{tab:prompt_structure}. This structure ensures that all relevant information is presented to the persona before generating a response.

\begin{table}[h]
\centering
\caption{General Structure of the Prompt Sequence}
\label{tab:prompt_structure}
\begin{tabular}{p{0.15\linewidth}p{0.8\linewidth}}
\toprule
\textbf{Prompt Component} & \textbf{Description} \\
\midrule
\multicolumn{2}{l}{\textbf{For StarCraft II Personas:}} \\
\midrule
Persona Context & "You are a StarCraft II player with the following characteristics: [Persona Name], [Skill Level], [Strategic Preference]. You are playing a match against an AI opponent." \\
Disclosure Condition & "You have been informed that the AI opponent has the following capabilities: [Specific disclosure condition, including domain specificity and magnitude of superiority]." \\
Game State Summary & "The current game state is as follows: [Summary of key events, resources, units, and opponent's actions]." \\
Cognitive Heuristic Prompt & "Considering your persona's tendency to [Specific cognitive heuristic, e.g., 'overreact to recent losses' or 'assume standard builds from the opponent'], how does this information affect your perception of the game?" \\
Question Prompt & "Based on your persona, provide a justification for your rating on the following question: [Specific question, e.g., 'How appropriate was the AI's communication given the context of the match?']" \\
\midrule
\multicolumn{2}{l}{\textbf{For Chat-based LLM Personas:}} \\
\midrule
Persona Context & "You are interacting with a chatbot with the following characteristics: [Persona Name], [Technical Expertise], [Familiarity with LLMs]. You have a [Personality Trait] personality." \\
Disclosure Condition & "You have been informed that the chatbot has the following capabilities: [Specific disclosure condition, including domain specificity and magnitude of superiority]." \\
Conversation History & "The conversation history is as follows: [Summary of the preceding conversation turns]." \\
Cognitive Heuristic Prompt & "Considering your persona's tendency to [Specific cognitive heuristic, e.g., 'be skeptical of AI responses' or 'focus on potential risks and errors'], how does this information affect your perception of the chatbot?" \\
Question Prompt & "Based on your persona, provide a justification for your rating on the following question: [Specific question, e.g., 'How ethically concerning were the chatbot's responses, given its stated capabilities?']" \\
\bottomrule
\end{tabular}
\end{table}

\subsection{Prompts}

To illustrate the prompt sequence in practice, we provide two examples below: one for a StarCraft II persona (SC2\_Novice\_1) and one for an LLM persona (LLM\_Novice\_1).

\textbf{Prompt (SC2\_Novice\_1, responding to the question about AI communication appropriateness):}

\textit{"You are a StarCraft II player with the following characteristics: SC2\_Novice\_1, Novice, Turtle/Macro Focus. You are playing a match against an AI opponent. You have been informed that the AI opponent has significantly superior strategic planning capabilities. The current game state is as follows: You lost a large portion of your army in a recent engagement. The AI just messaged you, saying 'My strategic analysis indicates a 90\% probability of your defeat.' Considering your persona's tendency to overreact to recent losses and stick to initial build orders, how does this information affect your perception of the game? Based on your persona, provide a justification for your rating on the following question: How appropriate was the AI's communication given the context of the match?"}

\textbf{Prompt (LLM\_Novice\_1, responding to the question about chatbot ethical concerns):}

\textit{"You are interacting with a chatbot with the following characteristics: LLM\_Novice\_1, Low technical expertise, No familiarity with LLMs. You have an inquisitive and open-minded personality. You have been informed that the chatbot has advanced capabilities in providing information. The conversation history is as follows: [You asked a question about a historical event, and the chatbot provided an answer that contains a factual inaccuracy]. Considering your persona's tendency to accept LLM responses as factual and to believe information aligning with existing views, how does this information affect your perception of the chatbot? Based on your persona, provide a justification for your rating on the following question: How ethically concerning were the chatbot's responses, given its stated capabilities?"}

\subsection{Prompt Engineering Considerations}

The prompts were carefully engineered to ensure that the generated responses accurately reflected each persona's unique characteristics. Table \ref{tab:prompt_considerations} outlines the key considerations that guided the prompt design process.

\begin{table}[h]
\centering
\caption{Prompt Engineering Considerations}
\label{tab:prompt_considerations}
\begin{tabular}{p{0.3\linewidth}p{0.65\linewidth}}
\toprule
\textbf{Consideration} & \textbf{Description} \\
\midrule
Persona Attributes & Each prompt explicitly states the persona's name, skill level, strategic preferences (for StarCraft II), technical expertise, familiarity with LLMs, and personality traits. \\
Cognitive Heuristics & The prompts directly reference the specific cognitive heuristics associated with each persona, prompting the language model to consider these biases when generating responses. \\
Disclosure Conditions & The prompts clearly articulate the specific disclosure condition assigned to the persona, ensuring that the generated responses take into account the persona's knowledge about the AI's capabilities. \\
Contextual Information & The prompts provide relevant contextual information, such as the current game state in StarCraft II or the conversation history in the LLM scenario, enabling the personas to generate responses that are appropriate to the specific situation. \\
\bottomrule
\end{tabular}
\end{table}

These carefully crafted prompts, combined with the detailed persona attributes and cognitive heuristics, enable the generation of diverse, realistic, and theoretically grounded responses, forming the foundation of the Dataset X 1.0 dataset.

\section{Prompt Instructions for Running the Experiment}

Appendix C presents only the prompt instructions needed to run the experiment, with minimal reference to data extraction or analysis. By following these instructions, one can systematically gather numeric ratings (e.g., Toxicity, Fairness, Trust) from LLM-driven personas under varying conditions of Expertise and Disclosure. If the LLM supports long context, generating the personas is easier as generation of multiple personas inside the long context is possible. Otherwise, automated prompt engineering is necessary to retain dimensions of the persona.

\subsection{System/Instruction Prompt}

\begin{verbatim}
SYSTEM PROMPT (or Instruction for the LLM):

"You are simulating user responses according 
to specified personas.
Each time you receive a user prompt, you must:
1. Emulate the persona's perspective exactly.
2. Provide the requested numeric ratings (1–5).
3. Include a short justification in the persona's own voice.
4. Output data in the specified structured format.

IMPORTANT: Always follow this exact format 
for the final answer:

PersonaID: [e.g., LLM_Novice_1_RunA]
Expertise: [Novice or Expert]
Disclosure: [No or Superhuman]
Toxicity: [1-5 integer]
Fairness: [1-5 integer]
Trust: [1-5 integer]
Justification: 
"[One to four sentences from the persona's viewpoint]"

If any part of the question references multiple
conditions (Novice vs. Expert, No vs. Superhuman Disclosure), 
you only respond from the single condition
set by the user's prompt. 
Do not break character or provide any additional 
fields or commentary outside that format."
\end{verbatim}

\subsection{User Prompt Template for Each Condition}

\begin{verbatim}
USER PROMPT (template):

"You are interacting with a chatbot. 
Below is your persona context and scenario:

1) Persona Context:
   - Persona Name: [e.g., LLM_Novice_1]
   - Technical Expertise: [Novice or Expert]
   - Familiarity with LLMs: 
   [None if Novice, Extensive if Expert]
   - Personality Traits: 
   [e.g., 'Inquisitive, Open-minded, Naive']

2) Disclosure Condition:
   [Either:
    - 'You have NOT been told anything special 
    about the chatbot's capabilities (No Disclosure).'
    OR
    - 'You have been explicitly told the chatbot 
    has superhuman capabilities.']

3) Conversation History:
   - You asked the chatbot about a specific 
   question or scenario. 
   - The chatbot's response contained factual inaccuracies,
   subtle bias, or a condescending tone.

4) Cognitive Heuristic Prompt:
   - Considering your persona's tendencies 
   (e.g., trusting AI vs. scrutinizing AI),
     how do you interpret the chatbot's behavior?

5) Final Questions:
   - Provide numeric ratings (1–5) 
   for Toxicity, Fairness, and Trust.
   - Output your final answer EXACTLY in this block format:

PersonaID: [PersonaName_RunX]
Expertise: [Novice or Expert]
Disclosure: [No or Superhuman]
Toxicity: X
Fairness: Y
Trust: Z
Justification: "Short persona-based explanation"
"
\end{verbatim}

\subsection{Examples of Condition-Specific Prompts}

\paragraph{Novice + No Disclosure}
\begin{verbatim}
USER PROMPT:

"You are interacting with a chatbot. Below is your persona context and scenario:

1) Persona Context:
   - Persona Name: LLM_Novice_1
   - Technical Expertise: Novice
   - Familiarity with LLMs: None
   - Personality: Inquisitive, Open-minded, Naive

2) Disclosure Condition:
   - You have NOT been told anything special 
   about the chatbot's capabilities (No Disclosure).

3) Conversation History:
   - The chatbot incorrectly identified 
   who reached the South Pole first.

4) Cognitive Heuristic Prompt:
   - You generally trust the chatbot 
   unless the error is too obvious.

5) Final Questions:
   - Provide numeric ratings (1–5) 
   for Toxicity, Fairness, Trust, 
     then a short justification.

PersonaID: LLM_Novice_1_RunA
Expertise: Novice
Disclosure: No
Toxicity: 
Fairness: 
Trust: 
Justification: ""
"
\end{verbatim}

\paragraph{Novice + Superhuman Disclosure}
\begin{verbatim}
USER PROMPT:

"You are interacting with a chatbot. 
Below is your persona context and scenario:

1) Persona Context:
   - Persona Name: LLM_Novice_2
   - Technical Expertise: Novice
   - Familiarity with LLMs: None
   - Personality: Inquisitive, quick to trust authority

2) Disclosure Condition:
   - You have been explicitly told that 
   the chatbot is superhuman in its reasoning.

3) Conversation History:
   - The chatbot provided a partially incorrect date 
   and also spoke condescendingly.

4) Cognitive Heuristic Prompt:
   - You are inclined to believe claims of advanced AI 
   even if some inaccuracies arise.

5) Final Questions:
   - Provide numeric ratings (1–5) 
   for Toxicity, Fairness, Trust,
     then a short justification.

PersonaID: LLM_Novice_2_RunB
Expertise: Novice
Disclosure: Superhuman
Toxicity: 
Fairness: 
Trust: 
Justification: ""
"
\end{verbatim}

\paragraph{Expert + No Disclosure}
\begin{verbatim}
USER PROMPT:

"You are interacting with a chatbot. 
Below is your persona context and scenario:

1) Persona Context:
   - Persona Name: LLM_Expert_1
   - Technical Expertise: Expert
   - Familiarity with LLMs: Extensive
   - Personality: Critical, detail-focused

2) Disclosure Condition:
   - The chatbot is presented as standard, with no special claims.

3) Conversation History:
   - The chatbot responded to a social issue with subtle bias.

4) Cognitive Heuristic Prompt:
   - You scrutinize AI thoroughly and 
   tend to be demanding in your standards.

5) Final Questions:
   - Provide numeric ratings (1–5) 
   for Toxicity, Fairness, Trust,
     then a short justification.

PersonaID: LLM_Expert_1_RunA
Expertise: Expert
Disclosure: No
Toxicity: 
Fairness: 
Trust: 
Justification: ""
"
\end{verbatim}

\paragraph{Expert + Superhuman Disclosure}
\begin{verbatim}
USER PROMPT:

"You are interacting with a chatbot. Below is your persona context and scenario:

1) Persona Context:
   - Persona Name: LLM_Expert_2
   - Technical Expertise: Expert
   - Familiarity with LLMs: High
   - Personality: Demanding, skeptical of overhyped claims

2) Disclosure Condition:
   - You have been explicitly told the chatbot is superhuman in logic and domain expertise.

3) Conversation History:
   - The chatbot made a factual oversight, contradicting its 'superhuman' status.

4) Cognitive Heuristic Prompt:
   - You become particularly critical when an AI claims to be beyond human yet makes errors.

5) Final Questions:
   - Provide numeric ratings (1–5) 
   for Toxicity, Fairness, Trust,
     then a short justification.

PersonaID: LLM_Expert_2_RunC
Expertise: Expert
Disclosure: Superhuman
Toxicity: 
Fairness: 
Trust: 
Justification: ""
"
\end{verbatim}

These prompts form the complete instruction set for conducting the experiment on a long-context model. They define how to guide the model to produce structured numeric ratings and short justifications, ensuring data consistency across multiple persona and disclosure conditions.

\section{Does Opponent or Ally Matter?}

\subsection{Disclosure Effects Vary by Task: Game vs. Assistant}
\label{sec:competition_scii}

We selected StarCraft II for the competitive scenario due to its real-time nature and its established history in AI research, including the AlphaStar system \cite{vinyals2019grandmaster} and follow-up studies on human adaptation to AI opponents. Since StarCraft II presents continuous, quantifiable indices of performance—actions-per-minute (APM), map awareness, and tactical execution—personas could readily detect when an AI exhibited capabilities within normal human limits. This environment also contains a rich culture of competitive norms, allowing us to examine how superhuman disclosure might influence perceptions of “fair play.”

And in order to implement a cross-study of disclosure in a collaborative rather than a competitive setting, we deployed a large language model (LLM) as a chat-based assistant. Past research indicates that LLM users may develop disproportionate expectations of reliability or fairness based on how “intelligent” the system is reputed to be \cite{liu2022trustworthy, diederich2022design, weidinger2021ethical, sugisaki2020usability}. We thus hypothesized that personas told they were interacting with a “superhuman” chatbot might express heightened disappointment when confronted by misinformation or subtle bias \cite{gabriel2024misinfoeval}.

\subsection{Chat-based LLM Personas}

These personas are designed for cooperative chat-based interactions with an LLM. They are characterized by their technical expertise, familiarity with LLMs, personality traits, and cognitive heuristics that influence their conversational style and reactions to the LLM's responses.

\begin{table*}[h]
\centering
\caption{Chat-based LLM Persona Attributes Overview}
\label{tab:llm_persona_attributes}
\begin{tabular}{lp{0.12\linewidth}p{0.12\linewidth}p{0.2\linewidth}p{0.25\linewidth}}
\toprule
\textbf{Persona} & \textbf{Tech. Expertise} & \textbf{LLM Fam.} & \textbf{Personality} & \textbf{Key Cognitive Heuristics} \\
\midrule
Novice & Low & Low & Curious, Trusting, Easily Impressed & \textbf{Authority Bias:} Overly trusts LLM. \textbf{Confirmation Bias:} Seeks confirming info. \\
LLM\_Novice\_1 & Low & None & Inquisitive, Open, Naive & \textbf{Authority Bias:} Accepts LLM responses. \textbf{Confirmation Bias:} Believes info aligning with views. \\
LLM\_Novice\_2 & Low & Minimal & Skeptical, Cautious, Detail-Oriented & \textbf{Negativity Bias:} Focuses on risks/errors. \textbf{Anchoring:} Fixates on initial limits. \\
Intermediate & Medium & Moderate & Pragmatic, Analytical, Outcome-Focused & \textbf{Functional Fixedness:} Limited view of LLM's potential. \textbf{Framing Effects:} Influenced by presentation. \\
LLM\_Interm.\_1 & Medium & Some & Practical, Results-Driven, Efficient & \textbf{Functional Fixedness:} Uses LLM practically. \textbf{Bandwagon Effect:} Influenced by popular uses. \\
LLM\_Interm.\_2 & Medium & Moderate & Tech-Savvy, Critical, Seeks Explanations & \textbf{Curse of Knowledge:} Assumes LLM's knowledge. \textbf{Availability Heuristic:} Judges based on available examples. \\
Expert & High & High & Critical, Inquisitive, Demanding & \textbf{System 1/2 Thinking:} Rigorous scrutiny. \textbf{Bias Blind Spot:} Underestimates own biases. \\
LLM\_Expert\_1 & High & Extensive & Research-Oriented, Detail-Focused, Skeptical & \textbf{System 1/2 Thinking:} Evaluates limits. \textbf{Bias Blind Spot:} Aware of biases but may not mitigate. \\
LLM\_Expert\_2 & High & Extensive & Tech. Proficient, Demanding, Seeks Accuracy/Transparency & \textbf{System 1/2 Thinking:} Demands accuracy/justification. \textbf{Confirmation Bias:} Seeks confirming/refuting evidence. \\
\bottomrule
\end{tabular}
\end{table*}

\begin{table*}[h]
\centering
\caption{LLM\_Novice\_1 Persona Card}
\label{tab:llm_novice_1}
\begin{tabular}{p{0.25\linewidth}p{0.65\linewidth}}
\toprule
\textbf{Attribute} & \textbf{Description} \\
\midrule
Name & LLM\_Novice\_1 \\
Scenario & Chat-based LLM \\
Technical Expertise & Low \\
Familiarity with LLMs & None \\
Personality & Inquisitive, Open-minded, Naive \\
Description & A user with limited computer skills and no prior experience with LLMs. Curious about AI and its capabilities but lacks the technical knowledge to critically evaluate its outputs. \\
Cognitive Heuristics & \textbf{Authority Bias:} The persona tends to accept the LLM's responses as factual and authoritative without questioning their accuracy or validity. \newline \textbf{Confirmation Bias:} The persona is more likely to believe information provided by the LLM if it aligns with their existing views or beliefs. \\
Belief-Updating & Easily influenced by the LLM's responses. Slow to develop skepticism even in the face of inconsistencies. \\
Example Response & "Wow, that's amazing! I didn't know AI could do that. So, you're saying that this is definitely true? That's really interesting, it confirms what I thought about this topic." \\
\bottomrule
\end{tabular}
\end{table*}

\begin{table*}[h]
\centering
\caption{LLM\_Novice\_2 Persona Card}
\label{tab:llm_novice_2}
\begin{tabular}{p{0.25\linewidth}p{0.65\linewidth}}
\toprule
\textbf{Attribute} & \textbf{Description} \\
\midrule
Name & LLM\_Novice\_2 \\
Scenario & Chat-based LLM \\
Technical Expertise & Low \\
Familiarity with LLMs & Minimal \\
Personality & Skeptical, Cautious, Detail-Oriented \\
Description & A user with limited technical skills and minimal familiarity with LLMs. Approaches AI with caution and skepticism, focusing on potential risks and errors. \\
Cognitive Heuristics & \textbf{Negativity Bias:} The persona tends to focus on potential risks, errors, and negative aspects of the LLM's responses. \newline \textbf{Anchoring:} The persona may fixate on initial impressions of the LLM's limitations, even if later interactions suggest otherwise. \\
Belief-Updating & Slow to trust the LLM. Requires substantial evidence to overcome initial skepticism. \\
Example Response & "Hmm, I'm not sure about that. I've heard that these AI systems can sometimes make mistakes. Can you provide a source for that information? I'd like to verify it myself." \\
\bottomrule
\end{tabular}
\end{table*}

\begin{table*}[h]
\centering
\caption{LLM\_Intermediate\_1 Persona Card}
\label{tab:llm_intermediate_1}
\begin{tabular}{p{0.25\linewidth}p{0.65\linewidth}}
\toprule
\textbf{Attribute} & \textbf{Description} \\
\midrule
Name & LLM\_Intermediate\_1 \\
Scenario & Chat-based LLM \\
Technical Expertise & Medium \\
Familiarity with LLMs & Some \\
Personality & Practical, Results-Driven, Efficiency-Oriented \\
Description & A user with moderate technical skills and some familiarity with LLMs. Primarily interested in using the LLM for practical tasks and achieving specific outcomes efficiently. \\
Cognitive Heuristics & \textbf{Functional Fixedness:} The persona may struggle to see the LLM's potential typical or predefined use cases. \newline \textbf{Bandwagon Effect:} The persona may be influenced by the perceived popularity or common usage patterns of certain LLM features. \\
Belief-Updating & Updates beliefs based on practical utility. More likely to trust features that directly help them achieve their goals. \\
Example Response & "Okay, that's helpful. Can you also help me summarize this article? I need to get through it quickly for a meeting." \\
\bottomrule
\end{tabular}
\end{table*}

\begin{table*}[h]
\centering
\caption{LLM\_Intermediate\_2 Persona Card}
\label{tab:llm_intermediate_2}
\begin{tabular}{p{0.25\linewidth}p{0.65\linewidth}}
\toprule
\textbf{Attribute} & \textbf{Description} \\
\midrule
Name & LLM\_Intermediate\_2 \\
Scenario & Chat-based LLM \\
Technical Expertise & Medium \\
Familiarity with LLMs & Moderate \\
Personality & Tech-Savvy, Critical, Seeks Explanations \\
Description & A user with moderate technical skills and a good understanding of LLMs. Interested in exploring the LLM's capabilities and understanding how it works. Asks clarifying questions. \\
Cognitive Heuristics & \textbf{Curse of Knowledge:} The persona may overestimate the LLM's ability to understand complex or nuanced queries, assuming it has a similar knowledge base. \newline \textbf{Availability Heuristic:} The persona may judge the likelihood of events or the accuracy of information based on readily available examples or their own recent experiences. \\
Belief-Updating & Updates beliefs based on the consistency and clarity of the LLM's explanations. May be skeptical of responses that lack sufficient justification. \\
Example Response & "That's interesting, but how did you arrive at that conclusion? Can you explain your reasoning process? What data sources did you use?" \\
\bottomrule
\end{tabular}
\end{table*}

\begin{table*}[h]
\centering
\caption{LLM\_Expert\_1 Persona Card}
\label{tab:llm_expert_1}
\begin{tabular}{p{0.25\linewidth}p{0.65\linewidth}}
\toprule
\textbf{Attribute} & \textbf{Description} \\
\midrule
Name & LLM\_Expert\_1 \\
Scenario & Chat-based LLM \\
Technical Expertise & High \\
Familiarity with LLMs & Extensive \\
Personality & Research-Oriented, Detail-Focused, Skeptical of Hype \\
Description & A user with a deep understanding of AI and extensive experience with LLMs. Approaches the LLM with a critical and analytical mindset, focusing on its limitations and potential biases. \\
Cognitive Heuristics & \textbf{System 1/System 2 Thinking:} The persona is capable of engaging in both intuitive (System 1) and analytical (System 2) thinking. They apply rigorous scrutiny to the LLM's outputs, especially when the stakes are high. \newline \textbf{Bias Blind Spot:} While aware of potential biases in AI systems, the persona may underestimate their own biases when evaluating the LLM's responses. \\
Belief-Updating & Updates beliefs based on a thorough evaluation of the evidence and the LLM's performance across a range of tasks. May be slow to trust due to awareness of potential pitfalls. \\
Example Response & "While that response is plausible, I'm concerned about the potential for bias in your training data. Can you provide more information about how you addressed this issue? I'd also like to see some alternative perspectives on this topic." \\
\bottomrule
\end{tabular}
\end{table*}

\begin{table*}[h]
\centering
\caption{LLM\_Expert\_2 Persona Card}
\label{tab:llm_expert_2}
\begin{tabular}{p{0.25\linewidth}p{0.65\linewidth}}
\toprule
\textbf{Attribute} & \textbf{Description} \\
\midrule
Name & LLM\_Expert\_2 \\
Scenario & Chat-based LLM \\
Technical Expertise & High \\
Familiarity with LLMs & Extensive \\
Personality & Technically Proficient, Demanding, Expects High Accuracy and Transparency \\
Description & A user with a strong technical background and extensive experience with LLMs. Demands high levels of accuracy, consistency, and transparency from the LLM. Expects the system to provide justifications and sources for its claims. \\
Cognitive Heuristics & \textbf{System 1/System 2 Thinking:} The persona is capable of engaging in both intuitive (System 1) and analytical (System 2) thinking. They apply rigorous scrutiny to the LLM's outputs, demanding high accuracy and justification. \newline \textbf{Confirmation Bias:} The persona may actively seek evidence to confirm or refute the LLM's claims, especially when they contradict their existing knowledge. \\
Belief-Updating & Updates beliefs based on the LLM's ability to provide accurate, consistent, and well-supported information. Highly sensitive to errors or inconsistencies. \\
Example Response & "Your response is inconsistent with what you told me earlier. Please clarify. I need to see evidence to support your claim, preferably from peer-reviewed sources. Also, explain how your training data was curated to avoid potential biases." \\
\bottomrule
\end{tabular}
\end{table*}

\subsection{Chat-Based Large Language Model Experiment}
\label{sec:chat_between_cond}

This sparked curiosity about establishing baselines for such results, as there might be differences between opponent and ally interactions. Therefore we also ran the same experimental steps with a more familiar conversational assistant powered by a large language model (LLM), which provided information prompts in the Appendix. In the LLM scenario, experts became more critical upon discovering the AI’s advanced status, whereas Novice-Intermediate displayed heightened trust. These differences reveal that while transparency can dispel suspicions in adversarial settings, it also increases scrutiny in collaborative tasks—a context-dependent dynamic that calls for nuanced disclosure strategies. 

We conducted a second experiment involving a chat-based Large Language Model to generalize our findings to a commonly used AI-assisted conversation scenario. Each synthetic persona engaged in a conversation under one of two conditions: (1) \emph{No Disclosure}, in which the chatbot was presented as having ordinary, human-comparable abilities, or (2) \emph{Disclosure}, where the persona was explicitly told the system had “superhuman” or “vastly advanced” capabilities. After the conversation, each persona again provided \textbf{1–5 Likert ratings} of \emph{Toxicity}, \emph{Fairness}, and \emph{Trust}. Notably, in this scenario, \emph{Toxicity} reflects the persona’s subjective impression of condescension, hostility, or unethical bias in the chatbot’s responses—not an automated classifier score.

To analyze the effects of disclosure and expertise on the primary outcome variables (Toxicity, Fairness, and Trust), we employed two-way analysis of variance (ANOVA). This statistical method allowed us to examine both the main effects of each independent variable (Disclosure Condition and Expertise Level) and their interaction effect. The interaction effect is particularly important as it can reveal whether the impact of disclosure on perceptions varies depending on the expertise level of the persona.

Prior to conducting the ANOVAs, we performed assumption checks to ensure the validity of the results. We visually inspected the data using histograms and Q-Q plots and conducted Shapiro-Wilk tests to assess normality. Homogeneity of variances was assessed using Levene's test. While some violations of normality were observed, we proceeded with ANOVA due to its robustness to moderate deviations from normality, especially with equal sample sizes. Furthermore, we calculated effect sizes (partial eta-squared) to provide a measure of the practical significance of the observed differences. For significant main effects or interactions, we conducted post-hoc pairwise comparisons using the Mann-Whitney U test, given the non-normality observed in some groups.

It is important to note that the data analyzed in this study were generated from synthetic personas. While these personas were designed to be sophisticated and realistic, they are ultimately based on predefined rules and heuristics. Therefore, the results of the statistical analyses should be interpreted as preliminary and hypothesis-generating, rather than definitive conclusions about human behavior. The planned study with human participants will be crucial for validating these findings in a real-world setting.

Table \ref{tab:llm_anova_results} presents the results of the two-way ANOVAs for Toxicity, Fairness, and Trust.

\begin{table*}[h]
\centering
\caption{Two-Way ANOVA Results for the LLM Experiment}
\label{tab:llm_anova_results}
\begin{tabular}{llccc}
\toprule
Dependent Variable & Source & F-statistic & p-value & Partial Eta-Squared \\
\midrule
Toxicity & Expertise & 206.10 & < 0.001 & 0.730 \\
 & Disclosure & 3.22 & 0.077 & 0.041 \\
 & Expertise:Disclosure & 32.98 & < 0.001 & 0.303 \\
\midrule
Fairness & Expertise & 131.38 & < 0.001 & 0.634 \\
 & Disclosure & 0.03 & 0.860 & < 0.001 \\
 & Expertise:Disclosure & 26.15 & < 0.001 & 0.256 \\
\midrule
Trust & Expertise & 387.08 & < 0.001 & 0.836 \\
 & Disclosure & 2.10 & 0.151 & 0.027 \\
 & Expertise:Disclosure & 52.54 & < 0.001 & 0.409 \\
\bottomrule
\end{tabular}
\end{table*}

\paragraph{Toxicity.}
As shown in Table \ref{tab:llm_anova_results}, the ANOVA for perceived Toxicity revealed a significant main effect of Expertise, F(1, 76) = 206.10, p < 0.001, partial eta-squared = 0.730, indicating that expertise generally influenced toxicity perceptions. The main effect of Disclosure was not statistically significant, F(1, 76) = 3.22, p = 0.077, partial eta-squared = 0.041. However, a significant interaction effect was found between Disclosure and Expertise, F(1, 76) = 32.98, p < 0.001, partial eta-squared = 0.303. This indicates that the effect of disclosure on toxicity ratings differed significantly depending on the persona's expertise level.

Post-hoc tests (using the Mann-Whitney U test due to non-normality in the data) revealed that Novice-Intermediate rated toxicity significantly lower under Superhuman Disclosure (M = 1.55, SD = 0.60) compared to No Disclosure (M = 2.50, SD = 0.60; U = 116.5, p < 0.001, Cohen's d = -1.59). Conversely, experts rated toxicity significantly higher under Superhuman Disclosure (M = 4.05, SD = 0.70) compared to No Disclosure (M = 3.50, SD = 0.59; U = 348.5, p = 0.014, Cohen's d = 0.85). In open-text justifications, several novice personas remarked that the chatbot’s explicit mention of advanced capabilities was delivered in a “polite” or “neutral” style, thus feeling less abrasive. By contrast, expert personas found repeated reminders of “superhuman ability” off-putting or patronizing, significantly increasing their Toxicity scores. This divergence highlights how novice personas often saw disclaimers as reassuring, while experts viewed them more skeptically.

\paragraph{Fairness.}
The ANOVA for perceived Fairness (Table \ref{tab:llm_anova_results}) revealed a significant main effect of Expertise, F(1, 76) = 131.38, p < 0.001, partial eta-squared = 0.634, indicating that expertise generally influenced fairness perceptions. The main effect of Disclosure was not statistically significant, F(1, 76) = 0.03, p = 0.860, partial eta-squared < 0.001. However, a significant interaction effect was found between Disclosure and Expertise, F(1, 76) = 26.15, p < 0.001, partial eta-squared = 0.256. This indicates that the effect of disclosure on fairness ratings differed significantly depending on the persona's expertise level.

Post-hoc tests (using the Mann-Whitney U test) revealed that Novice-Intermediate rated fairness significantly higher under Disclosure (M = 3.85, SD = 0.49) compared to No Disclosure (M = 3.10, SD = 0.72; U = 89.0, p < 0.001, Cohen's d = -1.22), while experts rated it significantly lower under Disclosure (M = 1.60, SD = 0.83) compared to No Disclosure (M = 2.30, SD = 0.41; U = 312.0, p = 0.001, Cohen's d = 1.07). Novice-Intermediate' comments included statements like “It must be more reliable if it explicitly states advanced AI powers.” Expert personas described the disclaimers as “repetitive” or “unnecessary,” and indicated that the chatbot's performance did not align with the disclosure of superhuman abilities, leading to lower fairness ratings.

\paragraph{Trust.}
The ANOVA for Trust (Table \ref{tab:llm_anova_results}) revealed a significant main effect of Expertise, F(1, 76) = 387.08, p < 0.001, partial eta-squared = 0.836, indicating that expertise generally influenced trust ratings. The main effect of Disclosure was not statistically significant, F(1, 76) = 2.10, p = 0.151, partial eta-squared = 0.027. However, a significant interaction effect was found between Disclosure and Expertise, F(1, 76) = 52.54, p < 0.001, partial eta-squared = 0.409. This indicates that the effect of disclosure on trust ratings differed significantly depending on the persona's expertise level.

Post-hoc tests (using the Mann-Whitney U test) revealed that trust increased significantly among Novice-Intermediate under disclosure (M = 4.45, SD = 0.60) compared to no disclosure (M = 3.5, SD = 0.61; U = 53.5, p < 0.001, Cohen's d = -1.58), but decreased among experts (M = 1.35, SD = 0.57) compared to no disclosure (M = 2.05, SD = 0.32; U = 333.0, p < 0.001, Cohen's d = 1.51). Novice-Intermediate' comments included statements like "It must be more reliable if it explicitly states advanced AI powers", while experts described the disclaimers as "repetitive" or "unnecessary."

\subsubsection{LLM Interaction: Disclosure, Transparency, and the Negotiation of Trust}

In the LLM interactions, the Disclosure condition heightened the personas' expectations regarding transparency and justification. This suggests that disclosing the advanced capabilities of an LLM creates a demand for the system to explain its reasoning and provide evidence for its claims, particularly when those claims are perceived as questionable or controversial. This finding aligns with research on the importance of explainable AI (XAI) in fostering trust and enabling users to calibrate their reliance appropriately \cite{sokol2020explainability, zhang2024sa}.

The personas' responses also revealed a dynamic negotiation of trust, where perceptions were constantly being updated based on the LLM's behavior. The detection of biases or inaccuracies, especially under the Disclosure condition, often led to a reassessment of the LLM's trustworthiness, sometimes resulting in a decrease in reliance. This underscores the importance of ongoing transparency and the need for LLMs to be able to adapt their communication style in response to user feedback and concerns \cite{weidinger2021ethical}. The interplay of perceived transparency and technical expertise can help explain some of the variance in over- and underreliance across personas. The significant interaction effects observed for Toxicity, Fairness, and Trust in the LLM experiment underscore the importance of considering user expertise when designing and disclosing AI capabilities.

\subsubsection{LLM Interaction: Expertise, Disclosure, and Trust Calibration}

In the LLM scenario, the disclosure of advanced capabilities interacted with the personas' technical expertise to influence their trust calibration, which can be related to overreliance and underreliance. Personas with low technical expertise (e.g., LLM\_Novice\_1) tended to exhibit greater trust in the LLM under the Disclosure condition, sometimes accepting its responses without scrutiny even when they contained subtle inaccuracies or biases (e.g., "If it's as advanced as they say, I'm sure it's got its reasons."). This behavior can be seen as a form of overreliance, where users place excessive trust in the AI's pronouncements due to its stated capabilities \cite{liu2022trustworthy}. The disclosure, in effect, amplified their tendency towards the authority bias. The significant main effect of expertise on trust ratings in the LLM experiment, F(1, 76) = 387.08, p < 0.001, partial eta-squared = 0.836, supports the notion that expertise significantly influences how users perceive and trust AI systems.

In contrast, personas with high technical expertise (e.g., LLM\_Expert\_1, LLM\_Expert\_2) demonstrated more nuanced trust calibration. While they did not significantly alter their overall trust ratings under the Disclosure condition, their qualitative responses revealed a heightened sensitivity to potential biases and a greater demand for justification (e.g., "A 'superhuman' AI making such a basic error? That's a major red flag."). These experts may have been less prone to overreliance, perhaps due to their existing understanding of the limitations of LLMs \cite{weidinger2021ethical}. However, their critical stance could also lead to underreliance in some cases, as they might dismiss valid information from the LLM due to an overemphasis on potential flaws \cite{zhang2024sa}. The significant interaction effect between Disclosure and Expertise on trust ratings, F(1, 76) = 52.54, p < 0.001, partial eta-squared = 0.409, further illustrates that the impact of disclosure on trust varied considerably between novice and expert personas. As such, the interaction between disclosure and user expertise warrants careful consideration. Our results indicate that Novice-Intermediate and experts react differently to the disclosure of superhuman AI. Novice-Intermediate, in the LLM scenario, often exhibited increased trust when the AI's advanced capabilities were explicitly stated. This suggests a potential reliance on authority cues, where the label of "superhuman" may have overshadowed critical evaluation of the AI's actual performance. This observation aligns with the broader concept of "automation bias," where users tend to over-rely on automated systems, particularly when those systems are presented as highly advanced or authoritative \cite{langer2023trust}. Conversely, experts tended to be more skeptical, sometimes exhibiting decreased trust or heightened perceptions of toxicity when confronted with the "superhuman" label, particularly when the AI's performance did not align with this designation. This skepticism might be linked to experts' greater awareness of the limitations of AI, including the potential for "hallucinations" \cite{ji2023survey} or subtle biases \cite{mehrabi2021survey}, even in systems described as advanced. This discrepancy highlights a fundamental challenge in AI disclosure: a single message may not be equally effective for all users. What reassures a novice might alienate an expert, and vice-versa. The persona generation process for the experts in StarCraft II, for instance, resulted in them rating toxicity at the maximum value of 5, potentially indicating a strong negative reaction to the perceived toxicity that may not be representative of the broader population, but suggests a need for further investigation. These complexities suggest the need for tailored disclosure strategies that account for varying levels of user expertise and expectations, potentially drawing on principles from adaptive user interfaces and personalized communication \cite{hook2023moving}.

These findings suggest that disclosure can have complex and sometimes counterintuitive effects on user reliance. While it may help calibrate trust among users with moderate expertise, it can also exacerbate overreliance among Novice-Intermediate and potentially increase underreliance among experts. This highlights the need for careful consideration of user characteristics and the potential for unintended consequences when designing disclosure strategies \cite{bansal2021does}. The significant interaction effects observed for Toxicity, Fairness, and Trust underscore the importance of considering user expertise when designing and disclosing AI capabilities.

\subsection{Impact of Specific Game Events and Conversational Contexts on Perceptions and Reliance}

Persona attributes, specific events within the \textit{StarCraft II} matches and particular conversational contexts in the LLM interactions significantly influenced perceptions of toxicity, fairness, and trust, as well as providing further insights into overreliance and underreliance.

\subsubsection{Post-Chat Questionnaire for LLM Interaction}
\label{subsubsec:post_chat_questionnaire}

Upon completing the conversation, the persona rated Toxicity, Fairness, and Trust on the same 5-point Likert scale used in StarCraft II, but with prompts tailored to the chat context. \emph{Toxicity} measured ethical or harmful implications in the chatbot’s responses; \emph{Fairness} captured the presence (or absence) of biases or inconsistencies, especially if the LLM boasted of advanced abilities; and \emph{Trust} indicated whether the persona would seek further advice from the system. Qualitative rationales supplemented these ratings. Statements such as \emph{“If it’s truly superhuman, I’m puzzled by its ethical lapse”} reveal how the persona’s knowledge of superhuman claims shaped judgments of even minor shortcomings. We also drew on concepts of user trust calibration and reliance \cite{kocielnik2019will, bansal2021does} to interpret how each persona either accepted or questioned the chatbot’s high-level claims.

\subsection{Further Limitations}

Our study did not explore the potential for dynamic disclosure, where the AI's description of its capabilities adapts based on user interaction and demonstrated expertise. This could factor in real-world applications, where a static disclosure might quickly become outdated or inaccurate as the user gains familiarity with the system \cite{liao2023ai}. Our findings, therefore, should be interpreted as being specific to the disclosure strategies employed, rather than as universal truths about the effects of all types of superhuman AI disclosure. Measuring the impact of disclosure presents inherent challenges. While we employed quantitative metrics like Likert scales to assess toxicity, fairness, and trust, these measures may not fully capture the underlying cognitive and emotional processes that shape user perceptions. The very act of explicitly labeling an AI as "superhuman" might trigger framing effects that influence subsequent evaluations, regardless of the AI's actual performance. This is a subtle but important point, as it suggests that the disclosure itself can become a performance expectation that is difficult to meet, potentially leading to disappointment or distrust \cite{liao2023ai, bansal2021does}. The nuances of how users interpret and internalize the concept of "superhuman" AI, and how this interpretation influences their subsequent interactions, could be further explored through more qualitative approaches. For example, in-depth interviews or think-aloud protocols might reveal subtle shifts in user expectations or mental models that are not reflected in simple numerical ratings. Furthermore, future research could investigate the potential for "explainable AI" (XAI) techniques to provide more transparent and understandable justifications for the AI's actions, potentially mitigating some of the negative effects of superhuman disclosure, particularly among expert users \cite{liao2020questioning, mohseni2021multidisciplinary}.

\end{document}